\documentclass{article}

\usepackage{graphicx}
\usepackage{natbib}

\usepackage{amssymb}
\usepackage{array,amsmath,amssymb,capt-of,ifthen,calc,graphicx,bm,soul,color,amstext}
\usepackage{graphicx}
\usepackage{esint}

\usepackage{tikz}

\newcommand{\lyxaddress}[1]{
\par {\raggedright #1
\vspace{1.4em}
\noindent\par}
}

\usepackage[a4paper,top=2.5cm,bottom=2.5cm]{geometry}
\date{}

\title{Lattice ellipsoidal statistical BGK model for\\ thermal non-equilibrium flows}

\author{Jianping Meng$^1$, Yonghao Zhang$^1$, Nicolas G. Hadjiconstantinou$^2$, \\Gregg A. Radtke$^2$ and Xiaowen Shan$^3$}
\begin{document}
\maketitle
\lyxaddress{$^{1}$Department of Mechanical \& Aerospace Engineering, University of Strathclyde, Glasgow G1 1XJ, UK}
\lyxaddress{$^{2}$Department of Mechanical Engineering,
  Massachusetts Institute of Technology, Cambridge,
  Massachusetts 02139, USA}
\lyxaddress{$^3$Exa Corporation, 55 Network Drive, Burlington, Massachusetts 01803, USA}

\begin{abstract}
A thermal lattice Boltzmann model is constructed on the basis of the ellipsoidal statistical Bhatnagar-Gross-Krook (ES-BGK) collision 
operator via the Hermite moment representation. The resulting lattice ES-BGK model uses a single 
distribution function and features an adjustable Prandtl number. Numerical simulations show that using a moderate 
discrete velocity set, this model can accurately recover steady and transient 
solutions of the ES-BGK equation in the slip-flow and early 
transition regimes in the small Mach number limit that is typical of microscale problems of practical interest. 
In the transition regime in particular, comparisons with numerical solutions of the ES-BGK model, 
direct Monte Carlo and low-variance deviational Monte Carlo simulations show good accuracy for values of the Knudsen 
number up to approximately $0.5$. 
On the other hand, highly non-equilibrium phenomena characterized by high Mach numbers, such as viscous heating and force-driven Poiseuille flow for large values of the driving force, are more difficult 
to capture quantitatively in the transition regime using discretizations chosen with computational efficiency 
in mind such as the one used here, although improved accuracy is observed as the number of discrete velocities is increased.
\end{abstract}

\section{Introduction}

The lattice Boltzmann (LB) method has been successful in simulating isothermal (athermal) flows; however, thermal flows still remain a challenge, despite significant efforts from a number of research groups~\citep[see][]{Shim2011,SBRAGAGLIA2009,scagliarini:055101, PhysRevE.79.066702,PhysRevE.78.016704, PhysRevE.79.066706,PhysRevE.76.036703, Zheng2010,PhysRevE.75.036704, 2007IJMPC..18..635S, PhysRevE.67.036306,PhysRevE.76.016702, PhysRevE.68.036706,1998AnRFM..30..329C}. Recent approaches fall into two broad categories: the high-order approach ~\citep[e.g.,][]{SBRAGAGLIA2009,2007IJMPC..18..635S,PhysRevE.67.036306}
and the double-distribution-function approach ~\citep[e.g.,][]{He1998282,PhysRevE.55.2780}. As a direct extension of the isothermal LB model, the high-order model only uses one distribution function while retaining high-order terms in the equilibrium distribution function to recover the full Navier-Stokes (NS) equations. Unfortunately, this approach requires  a richer velocity space  and as a result suffers in terms of simplicity and in some cases numerical instability\footnote{We also note some efforts on constructing single-distribution-function thermal models using standard lattices e.g. a 2 dimensional model with nine discrete velocities~\citep[see][]{PhysRevE.76.016702}. }.  
By contrast, in 
the double-distribution-function approach, two distribution functions are utilized: 
one for the velocity field and the other for the temperature field. 
As a result, high-order terms can be avoided and the standard
lattice can be used. In addition, this approach is numerically more stable. 
However, to utilize the standard lattice, the energy and momentum transport equation need to be decoupled through the Boussinesq approximation ~\citep[cf.][]{PhysRevE.75.036704}; in this case the double distribution approach 
does not recover the full NS equations in a strict sense. 

In addition to heat transfer, kinetic effects can make flows even more complex. For example, for gaseous
flows at the microscale, kinetic effects have to be taken into account as the Knudsen number 
(Kn, the ratio of the mean free path and the characteristic length) becomes finite~\citep{Hadjiconstantinou2006}. 
Due to its kinetic origins, the LB method has been shown to be able to describe moderately complex kinetic effects 
~\citep[see, e.g.,][]{zhang:047702,Toschi2005,sbragaglia:093602, Sbragaglia2006, tang:046701, zhang:046704, 2006JFM550413S, 
Ansumali2007, Kim20088655, yudistiawan:016705,guo:074903,Succi2002, kim:026704, tang:027701,Tian2007}.
Particularly, as the LB method can be considered as an approximation to the Boltzmann-BGK equation, 
high-order models should, in principle, be able to go beyond NS
hydrodynamics~\citep{2006JFM550413S,Meng2011a}.  Both analytical and numerical
analysis have shown that using an appropriate discrete velocity set, LB models 
can capture kinetic effects including
the Knudsen layer qualitatively and provide reasonably
accurate results for a range of isothermal problems~\citep{Kim20088655,yudistiawan:016705}; also, as expected, 
more discrete velocities generally give better 
predictions~\citep{Meng2011a}. When gas-surface interactions are concerned, 
discrete velocity sets from even-order 
Hermite polynomials typically perform better ~\citep{Meng2011b}. 
For thermal rarefied flows,
despite that the velocity-slip and temperature-jump problems 
have been investigated using a high-order LB model ~\citep{PhysRevE.79.066706}, 
significant effort is still required to develop robust LB methods 
for modeling general thermal non-equilibrium flows of engineering interest.

The Hermite expansion provides a systematic approach to derivation of the LB models for representing the NS hydrodynamic systems and beyond~\citep{2006JFM550413S}. The usage of the Hermite expansion was pioneered by \citet{Grad1958,12197}  for approximating solutions to the Boltzmann equation.  An important feature of Hermite polynomials is that the expansion coefficients correspond directly to moments of the distribution function. In addition, the truncation of higher-order terms does not directly alter the lower-order velocity moments of the distribution function. In his famous paper, Grad kept the first thirteen Hermite coefficients and obtained the well known 13-moment system ~\citep{12197}.  This system contains physics beyond the NS equation. However, a significant weakness of Grad's 13-moment system is its inability to describe smooth shock structures above a critical Mach number. Also, this system is not easily related a priori to the Knudsen number. As a result, significant effort has been devoted towards the development of an improved moment system in the last decade. Regularization for Grad's 13 moment system (i.e., the so-called R13 system) has been introduced on the basis of a Chapman-Enskog expansion around a non-equilibrium state ~\citep{struchtrup:2668,doi:10.1137/040603115,struchtrup:3921}. The related issues associated with the numerical scheme and boundary conditions have then been discussed by \citet{Torrilhon20081982} and \citet{Gu2007263}. A series of analytical solutions were also obtained and compared to the kinetic solutions \citep[see][]{taheri:017102,taheri:112004}. In addition to the R13 system, higher-order moment equation models have also been developed \citep{GU2009}.

Inspired by Grad's work, the Hermite expansion has
  been used to derive new LB models
  \citep{2006JFM550413S,PhysRevLett.80.65}. It was shown
  that the truncation of the Hermite expansion is equivalent
  to evaluating the distribution function at the chosen
  order of the discrete velocities. Thus, the resulting LB models can describe gaseous systems at the corresponding level. However, contrary to Grad's approach, the governing equation of LB models is presented in a much simpler form (e.g., the linear convective term)  as the evolution is accomplished at the distribution function level rather than the state-variable level. Therefore, the LB model has some advantages, e.g., straightforward numerical implementation and code parallelisation. 

In this work, we establish a thermal LB model
based on the ellipsoidal statistical Bhatnagar-Gross-Krook (ES-BGK)
equation following the systematic procedure of high-order Hermite expansion. The resulting model features an adjustable Prandtl
number in contrast to the BGK model whose Prandtl number is fixed at 1. 
We  validate our model using shear-driven (Couette) flows, Fourier flows and unsteady boundary heating problems 
in the small Mach-number limit over a range of Knudsen numbers. Using only a moderate set of discrete velocities,  
we find very good agreement with 
benchmark solutions. We also show that, with the same 
moderate discrete velocity set,  highly 
non-equilibrium problems characterized by high Mach numbers, such as force-driven 
Poiseuille flow for large values of the driving force, are captured accurately in the slip-flow regime and with good qualitative accuracy in the transition regime. Better accuracy requires a higher number of discrete velocities, in addition to higher-order Hermite expansion.
\section{Thermal lattice Boltzmann model}

\subsection{The ES-BGK equation}

The difficulty associated with solving the Boltzmann equation is mainly due to the collision term. To reduce the complexity, the collision term may be replaced by a simple statistical model that preserves 
the conservation laws of mass, momentum and energy. The most commonly-used model is known as the Bhatnagar-Gross-Krook (BGK) collision operator, which is often used in constructing lattice Boltzmann models. Despite its success, the BGK model suffers from a number of drawbacks: specifically, it yields a fixed Prandtl number (Pr) of unity, while the correct value for a monatomic gas is close to
$2/3$. To address this limitation, ~\citet{jr.:1658} proposed the ES-BGK model which replaces the Maxwellian distribution
of the standard BGK equation with an anisotropic Gaussian distribution. This model can be written as 
\begin{equation}
  \frac{\partial \hat{f}}{\partial \hat{t}}+\hat{c}_{i}\frac{\partial \hat{f}}{\partial \hat{x}_{i}}+\hat{g}_{i}\frac{\partial \hat{f}}{\partial \hat{c}_{i}}=-\frac{1}{\hat{\tau}}(\hat{f}-\hat{f}_{ES})\label{eq:esbgk}, 
\end{equation}
where $\hat{f}$ denotes the single-particle distribution function, $\hat{c}_{i}$ the
phase (particle) velocity, $\hat{g}_{i}$ the body force and $\hat{\tau}$  the mean
relaxation time which is assumed to be independent of particle velocity. The
anisotropic Gaussian distribution can be written as 


\begin{equation}
  \hat{f}_{ES}=\hat{\rho}\frac{1}{\sqrt{\det[2\pi\hat{\lambda}_{ij}]}}\exp\left[-\frac{1}{2}\hat{\lambda}_{ij}^{-1}\hat{C}_{i}\hat{C}_{j}\right],\label{eq:cfeq}\end{equation} 
where $\hat{\lambda}_{ij}=R\hat{T}\delta_{ij}+(b\hat{\sigma}_{ij})/ \hat{\rho}$ and $R$ is the gas constant. Macroscopic quantities, such as density $\hat{\rho}$, velocity $\hat{\bm{u}}$, shear stress $\hat{\bm {\sigma}}$, temperature $\hat{T}$ and heat flux $\hat{\bm {q}}$, can be obtained by integrating the
distribution function over the velocity space, i.e.
\begin{equation}
\label{eq:mac}
\left[\begin{array}{c}
\hat{\rho}\\
\hat{\rho} \hat{u}_{i}\\
\hat{\sigma}_{ij}\\
\hat{q}_{i}\\
\hat{\rho} DR\hat{T}
\end{array}\right]=\int \hat{f}\left[\begin{array}{c}
1\\
\hat{c}_{i}\\
\hat{C}_{<i}\hat{C}_{j>}\\
\frac{1}{2}\hat{C}_{i}\hat{C}_{j}\hat{C}_{j}\\
\hat{C}_{i}\hat{C}_{i}
\end{array}\right]d\hat{\bm{c}},
\end{equation}
where $\hat{C}_{i}=\hat{c}_{i}-\hat{u}_{i}$ is the particle
peculiar velocity and the angle bracket denotes the
trace-free tensors~\citep[see][Appendix A.2.2]{Struchtrup2005}. The pressure $\hat{p}$ can be related to the density and temperature by the equation of state 
\begin{equation}
\hat{p}=\hat{\rho} R\hat{T}.\label{eq:state}
\end{equation}
As $\hat{\lambda}_{ij}^{-1}$ (the inverse of the matrix) must be positive
definite, $b$ is restricted to $-\frac{1}{2}\leqslant b\leqslant1$. The exact values of $b$ and $\hat{\tau}$ can be determined by
the Boltzmann integral with a known inter-molecular force law or experimental data. In this work, experimental values will be used. The Navier Stokes equations can be derived
~\citep{jr.:1658} using a Chapman-Enskog expansion. The viscosity and thermal conduction coefficients can be written as $\hat{\mu}=(\hat{p}\hat{\tau})/(1-b)$ and $ \hat{\kappa}=\hat{p}R(D+2)\hat{\tau}/2$, respectively, where $D$ is the spatial dimension. Therefore, the Prandtl number is given by $Pr=1/(1-b)$ and can be adjusted via the free parameter
$b$. For thermal flows, viscosity depends on temperature, and can be expressed as $
\hat{\mu}/\mu_{0}=(\hat{T}/T_{0})^{\varpi}$, where $\varpi$ is related to the molecular interaction model,
varying from $0.5$ for hard-sphere molecular interactions to $1$
for Maxwellian interactions. Therefore, in general, $\hat{\tau}$ may be taken to be 
a function of temperature. 

The ES-BGK model has received renewed attention, in part because 
the associated  H theorem was recently proven
~\citep{Andries2000}. Several studies have shown that it 
can provide reasonable accuracy for modeling transport in simple  rarefied flows
~\citep{Andries2002,doi:10.1080/15567260701337209,Graur2009,Han2011,
mieussens:2797,Gallis2011,Garzo1995}.
In the following, we will develop a thermal LB model
based on the ES-BGK equation using the Hermite moment representation ~\citep{2006JFM550413S}; we will refer to the resulting LB method as the lattice ES-BGK model. In order to separate the numerical error arising from the LB method from the modeling error associated with approximating the hard-sphere operator with the ES-BGK model, we will also present {\it numerical} solutions of the Boltzmann equation with the ES-BGK collision operator obtained using the finite difference/discrete velocity method presented in ~\citet{PhysRevE.65.026315}. Although this method is not practical for problems of engineering interest due to the high dimensionality associated with the very fine discretization of the velocity space, it can be used to obtain accurate solutions of the ES-BGK equation in the one-dimensional benchmark problems considered here. In order to differentiate from our lattice-based solutions, we will refer to the  ones obtained by the finite difference/discrete velocity method as {\it numerical}.

\subsection{Derivation of the lattice ES-BGK equation}

\citet{2006JFM550413S} have shown that a hierarchy of different order LB models can be
systematically derived from the BGK equation via the Hermite expansion, which can be regarded as approximations to the
BGK equation. Gas flows can then be described at the kinetic level
with various discrete velocity sets. This procedure can also be used to
derive the lattice ES-BGK model. For convenience, we use the following non-dimensional system,
\[
x_{i}=\frac{\hat{x}_{i}}{L},u_{i}=\frac{\hat{u}_{i}}{\sqrt{RT_{0}}},t=\frac{\sqrt{RT_{0}}\hat{t}}{L},g_{i}=\frac{L\hat{g}_{i}}{RT_{0}},c_{i}=\frac{\hat{c}_{i}}{\sqrt{RT_{0}}},T=\frac{\hat{T}}{T_{0}},
\]

\[
\tau=\frac{\sqrt{RT_0}\hat{\tau}}{L},f=\frac{\hat{f}(RT_{0})^{D/2}}{\rho_{0}},\rho=\frac{\hat{\rho}}{\rho_{0}},p=\frac{\hat{p}}{p_{0}},\mu=\frac{\hat{\mu}}{\mu_{0}},q_{i}=\frac{\hat{q}_{i}}{p_{0}\sqrt{RT_{0}}},\sigma_{ij}=\frac{\hat{\sigma}_{ij}}{p_{0}}.
\]
The temperature and state equations become 
\[
DT=\frac{1}{\rho}\int f C_{i} C_{i}d{\bm{c}},
\]
\[
p=\rho T.
\]
The non-dimensional
relaxation time can be written explicitly as 
\[
\tau=\frac{\mu_0 \sqrt{RT_0}}{p_0L}\frac{\mu}{Pr p}=\frac{Kn}{Pr}\frac{\mu}{p},
\]
where the Knudsen number is defined by 
\[
Kn=\frac{\mu_{0}\sqrt{RT_{0}}}{p_{0}L}.
\]

To solve Eq.(\ref{eq:esbgk}), the distribution function 
is first projected onto a functional
space spanned by the Hermite basis. If a Gauss-Hermite
quadrature of a degree $\geqslant 2N$ is chosen, then the
first $N$ velocity moments of the distribution function are
retained \citep[see][P.420]{2006JFM550413S}. Therefore, the
distribution function is essentially  approximated as
\begin{equation}
  f(\bm{x},\bm{c},t)\approx \omega(\bm{c})\sum_{n=0}^{N}\frac{1}{n!}\bm{a}^{(n)}(\bm{x},t)\colon\bm{\chi}^{(n)}(\bm{c}),\label{approxf}
\end{equation} and  the coefficients $\bm{a}^{(n)}$ are calculated from 
\begin{equation}
  \bm{a}^{(n)} =  \int f\bm{\chi}^{(n)}d\bm{c} 
  \label{an} 
\end{equation}
where $\bm{\chi}^{(n)}$ is the $n$-th order Hermite polynomial and $\omega(\mathbf{c})$
is the weight function.

The key step is to expand the anisotropic Gaussian distribution as
\begin{equation}
  f_{ES}\approx f_{ES}^N=\omega(\bm{c})\sum_{n=0}^{N}\frac{1}{n!}\bm{a}_{ES}^{(n)}\colon\bm{\chi}^{(n)}(\bm{c}),\label{approxfeq}
\end{equation} where
\begin{equation}
  \bm{a}_{ES}^{(n)}=\int f_{ES}\bm{\chi}^{(n)}d\bm{c}
  \approx \sum_{\alpha=1}^{d}\frac{w_\alpha}{\omega(\bm
    c_\alpha)}f_{ES}^{N} \bm \chi^{n}(\bm c_\alpha),
\end{equation} 
and $w_{\alpha}$ and $\bm{c}_{\alpha}$, $\alpha=1,\cdots,d$, are the
weights and abscissae of a Gauss-Hermite quadrature. For the
anisotropic distribution, the
first a few coefficients are given by 
\begin{equation}
a_{ES}^{(0)}=\rho,
\end{equation}
\begin{equation}
a_{ES,i}^{(1)}=\rho u_{i},
\end{equation}
\begin{equation}
a_{ES,ij}^{(2)}=\rho(u_{i}u_{j}+\lambda_{ij}-\delta_{ij}),
\end{equation}
\begin{equation}
a_{ES,ijk}^{(3)}=\rho(u_{i}u_{j}u_{k}+\lambda_{ij}u_{k}+\lambda_{ik}u_{j}+\lambda_{jk}u_{i}-\delta_{ij}u_{k}-\delta_{ik}u_{j}-\delta_{jk}u_{i}),
\end{equation}
\begin{eqnarray}
a_{ES,ijkl}^{(4)} & = & \rho[u_{i}u_{j}u_{l}u_{k}+(\lambda_{il}-\delta_{il})u_{j}u_{k}+(\lambda_{ij}-\delta_{ij})u_{l}u_{k}+(\lambda_{ik}-\delta_{ik})u_{j}u_{l} \nonumber \\
 & + & (\lambda_{jl}-\delta_{jl})u_{i}u_{k}+(\lambda_{jk}-\delta_{jk})u_{i}u_{l}+(\lambda_{kl}-\delta_{kl})u_{i}u_{j} \\
 & + & (\lambda_{ij}-\delta_{ij})(\lambda_{kl}-\delta_{kl})+(\lambda_{ik}-\delta_{ik})(\lambda_{jl}-\delta_{jl})+(\lambda_{il}-\delta_{il})(\lambda_{kj}-\delta_{kj})] \nonumber.
 \end{eqnarray}
The body force term $F(\bm{x},\bm{c},t)=-\bm{g}\cdot\nabla_{c}f$
can be approximated as:
\begin{equation}
  F(\bm{x},\bm{c},t)=\omega\sum_{n=1}^{N}\frac{1}{(n-1)!}\bm{g}\bm{a}^{(n-1)}\colon\bm{\chi}^{(n)}.\end{equation}

Although Eq.~(\ref{eq:cfeq}) has an infinite Hermite expansion, a LB model is built under the assumption that to some level of approximation only the leading terms contribute explicitly to the hydrodynamics \citep[see][]{2006JFM550413S} and thus the infinite series can be truncated. In this work, a 4th-order Hermite expansion will be used. 

Given the above discussion, the ES-BGK equation Eq.(\ref{eq:esbgk}) can be rewritten in its truncated form, i.e., 
\begin{equation}
\frac{\partial f}{\partial t}+c_i \frac{\partial f}{\partial x_i}=-\frac{1}{\tau}(f-f^{N}_{ES})+F.
\label{truncatedeq}
\end{equation}  

For a 4th-order expansion, a discrete velocity set with
9th-order or higher accuracy is required ~\citep{2006JFM550413S}.  
There are a number of ways of constructing a high-order discrete velocity set 
~\citep{2006JFM550413S,PhysRevE.81.036702,PhysRevE.79.046701,Chikatamarla2006}.
A straightforward approach is to utilize the roots of the Hermite polynomial. In
one dimension, the discrete velocities $c_{\alpha}$ are just the roots of
the Hermite polynomials, and the corresponding weights are determined by: 
\begin{equation}
w_{\alpha}=\frac{n!}{[n\chi^{n-1}(c_{\alpha})]^{2}}.\label{weight}
\end{equation} 
Another useful procedure is to utilize the entropy
construction \citep[cf.][]{Chikatamarla2006,PhysRevE.79.046701}. Given 
one-dimensional velocity sets, those of the higher-dimension can
be constructed using the \lq\lq production\rq\rq~formulae
~\citep{2006JFM550413S}.  Once the discrete velocity set is chosen, the governing
equation of the lattice ES-BGK model can be written as 
\begin{equation}
  \frac{\partial f_{\alpha}}{\partial t}+c_{\alpha,i}\frac{\partial
    f_{\alpha}}{\partial
    x_{i}}=-\frac{1}{\tau}\left(f_{\alpha}-f_{ES,\alpha}\right)+g_{\alpha},
  \label{lbgk}
\end{equation}
where
$f_{\alpha}=\frac{w_{\alpha}f(\bm{x},\bm{c}_{\alpha},t)}{\omega(\bm{c}_{\alpha})}$,
$f_{ES,\alpha}=\frac{w_{\alpha}f_{ES}(\bm{x},\bm{c}_{\alpha},t)}{\omega(\bm{c}_{\alpha})}$
and
$g_{\alpha}=\frac{w_{\alpha}F(\bm{x},\bm{c}_{\alpha},t)}{\omega(\bm{c}_{\alpha})
}$.

\subsection{Remarks on the  accuracy  beyond Navier-Stokes}
\label{remarkaccuracy}
As the distribution functions $f$ and $f_{ES}$ are approximated and evaluated at the Gauss-Hermite quadrature points,
cf Eq.(\ref{approxf}), we can relate the chosen
Gauss-Hermite quadrature to the approximated moments of the distribution function. 
Generally speaking, more discrete velocities mean higher-order moments
can be obtained more accurately. With sufficiently accurate
quadrature and retaining increasingly high-order terms of the expanded $f_{ES}$,
the LB model is becoming a better approximation of the original ES-BGK equation (cf Fig.\ref{cvg}). 

When the Chapman-Enskog expansion is valid, we can evaluate model accuracy in terms of
the Knudsen number. According to the Chapman-Enskog expansion,  the distribution function $f$ is assumed to be given by an
asymptotic series expanded in powers of the (formal) small parameter $\epsilon$ \citep[see][Chapter 4]{Struchtrup2005},
\begin{equation}
f=f^{0}+\epsilon f^{(1)}+\epsilon^2 f^{(2)} + \cdots 
\end{equation} where $\epsilon$ plays the role of the
Knudsen number; $f^0$ is the Hermite approximation of the  Maxwellian $f_{M}$ (its first four Hermite coefficients can be found in \citet{2006JFM550413S}, cf. Eq.(3.12))). Furthermore, the time and spatial variations are also measured in powers of $\epsilon$, i.e., 
\begin{equation}
\partial_t=\epsilon \partial_t^{(0)}+\epsilon^2 \partial_t^{(1)}+ \cdots
\end{equation}  and $\bm \nabla=\epsilon \bm \nabla$ \citep{2006JFM550413S}.
Considering the form of $f^0$, $f^{N}_{ES}$ can be written as
\begin{equation}
f^{N}_{ES}=f^{0}+\omega\left[\frac{1}{2} b \sigma_{ij} \chi^{(2)}_{ij}+\frac{1}{6}(\sigma_{ij} u_k +\sigma_{ik} u_j+\sigma_{jk} u_i) \chi_{ijk}^{(3)}+\cdots \right].
\end{equation}
The shear stress $\sigma_{ij}$ should be also expanded as 
\begin{equation}
\sigma_{ij}=\epsilon \sigma_{ij}^{(1)}+\epsilon^2 \sigma_{ij}^{(2)}+\cdots,
\end{equation} 
and $f^{N}_{ES}$ can thereby be written as,
\begin{equation}
f^{N}_{ES}=f^{0}+\epsilon f_{ES}^{N,(1)}+\epsilon^2 f_{ES}^{N,(2)}+\cdots,
\end{equation} where 
\begin{equation}
f_{ES}^{N,(1)}=\omega\left[\frac{1}{2} b \sigma_{ij}^{(1)} \chi^{(2)}_{ij}+\frac{1}{6}(\sigma_{ij}^{(1)} u_k +\sigma_{ik}^{(1)} u_j+\sigma_{jk}^{(1)} u_i) \chi_{ijk}^{(3)}+\cdots \right],
\end{equation} and 
\begin{equation}
f_{ES}^{N,(2)}=\omega\left[\frac{1}{2} b \sigma_{ij}^{(2)} \chi^{(2)}_{ij}+\frac{1}{6}(\sigma_{ij}^{(2)} u_k +\sigma_{ik}^{(2)} u_j+\sigma_{jk}^{(2)} u_i) \chi_{ijk}^{(3)}+\cdots \right].
\end{equation}
By matching the terms in the same powers of $\epsilon$, we
can have the solution for the two leading orders, namely the NS order
\begin{equation}
f^{(1)}=f_{ES}^{N,(1)}-\tau(\partial_t^{(0)}+\bm c \cdot \bm \nabla + \bm g \cdot \bm \nabla_c )f^{(0)}
\end{equation} and the Burnett order 
\begin{equation}
f^{(2)}=f_{ES}^{N,(2)}-\tau[(\partial_t^{(0)}+\bm c \cdot \bm \nabla + \bm g \cdot \bm \nabla_c )f^{(1)}+\partial_t^{1}f^{0}].
\end{equation}
 It is
rather tedious to calculate the explicit form of
$f^{(1)}$ and $f^{(2)}$. However, here we only need to know their highest order as a Hermite polynomial of
the particle velocity $\bm c$. To determine the highest-order, we
notice  that the operators $\partial_t^{0}$ and $\partial_t^{1}$  do not alter the
order while the operators $\bm c \cdot \bm \nabla$ and $\bm \nabla_c$ 
 increase the order by one. Therefore, the Burnett order solution $f^{(2)}$ is a polynomial of
$\bm c$ of two-order higher than $f_{ES}^{N}$.  If 
desired, this discussion can be extended to solutions of higer order.  Here we restrict the discussion to the
first two orders.

With the above discussion, we can estimate the accuracy level associated with the
chosen expansion order and quadrature. As 
fourth-order terms of $\bm c$ are retained in the expansion of $f_{ES}$, the
Burnett-order solution of the distribution function can be represented by a
sixth-order polynomial. For the heat flux, a third-order
velocity moment, to be calculated with the Burnett level of
accuracy, the chosen quadrature 
has to be accurate for an integration over the full space
for a ninth-order polynomial. Hence, according to the Gaussian
quadrature rule \citep[cf.,][Appendix in]{2006JFM550413S},  it is necessary to adopt a quadrature with
an algebraic degree of precision beyond the ninth-order.
However, as the ninth-order quadrature is not optimal for 
describing gas-wall interactions~\citep{Meng2011b}, an
11th-order quadrature is chosen in this work. 

It is worth noting again that the above analysis relies on the
validity of the Chapman-Enskog expansion. Consequently, one
should be careful when interpreting the above conclusion for large Knudsen numbers. Furthermore, it may
be better to understand the requirement on the quadrature
for the corresponding accuracy as necessary rather than sufficient. Nevertheless, the above discussion is 
useful for understanding the capability of the LB model,
which can be seen in numerical simulations below.

\section{Numerical implementation}
\subsection{Scheme}
\begin{figure}
\begin{centering}
\includegraphics[width=0.48 \textwidth]{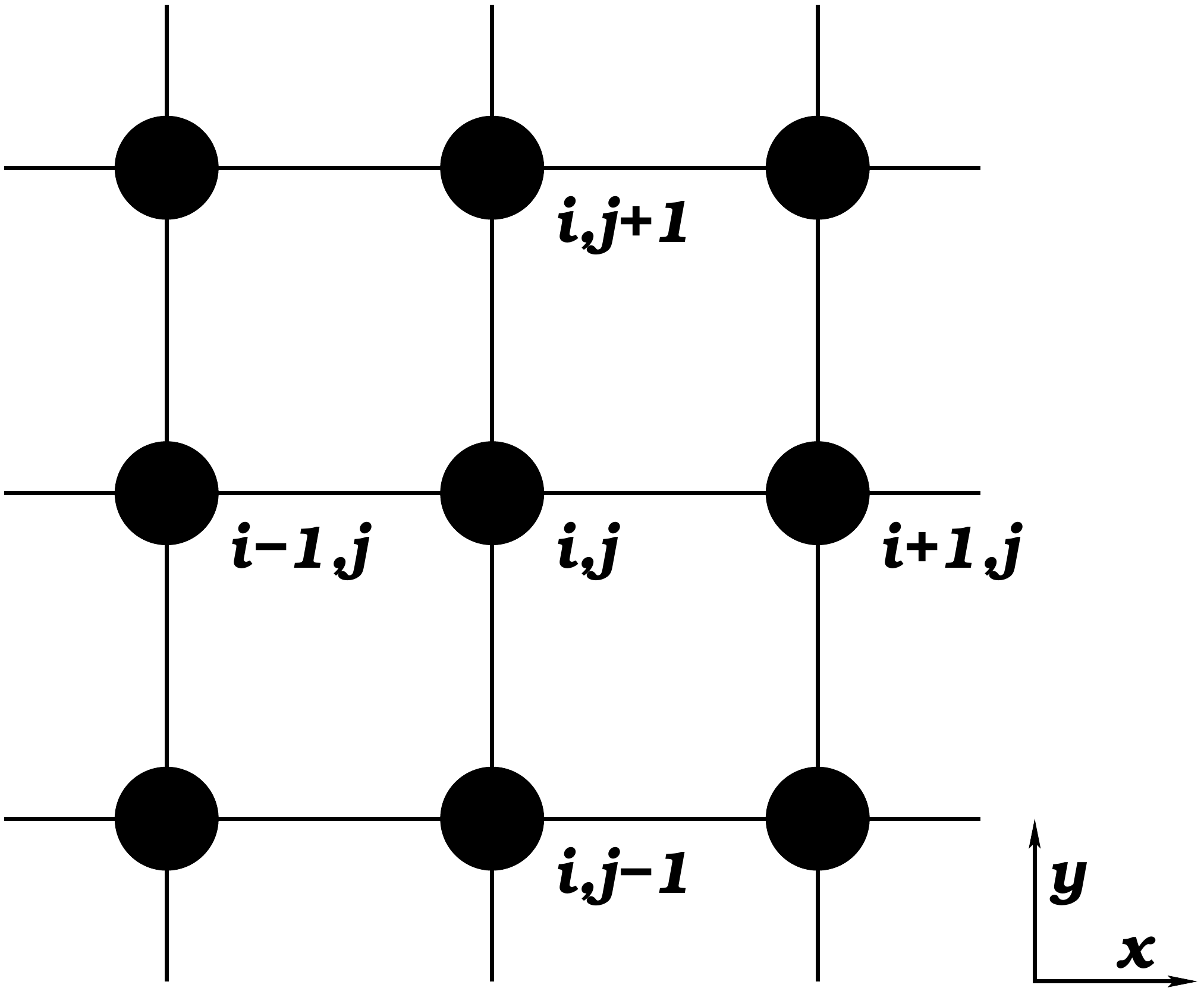} \caption{Schematic diagram of square lattices.\label{figgrid} }
\end{centering}
\end{figure}
To solve Eq.~(\ref{lbgk}), various numerical schemes can be used, depending
on the characteristics of the problem of interest.
Typically, if a first-order upwind finite difference is
chosen, we have a standard stream-collision scheme. 

As flows with various Knudsen numbers will
be considered in the following simulations, the numerical scheme has to be
carefully chosen to cope with discontinuities at the system
boundaries caused by rarefaction ~\citep{Hadjiconstantinou2006}. In particular,
in order to intrinsically describe the gas-surface interaction, the kinetic
boundary condition should be accurately implemented in the numerical scheme.
Meanwhile, there is still
no general and accurate implementation of the
standard stream-collision scheme for high-order models, especially for rarefied flows. On the other hand, for high-order LB models, the discrete velocity points may not
coincide with the lattice points, which makes the standard stream-collision procedure impossible. Therefore, in the interest of simplicity, a  
finite difference formulation has been chosen and more specifically, a 1st-order forward Euler method is coupled to a 2nd-order total variation diminishing (TVD) scheme
~\citep{Kim20088655,Sofonea2009,Toro2009}.  

The resulting scheme can be summarized  as
follows. Let $f_{\alpha,i}^{n,j}$ denote the distribution function value $f_{\alpha}$ at the node ($x_{i}$, $y_{j}$) at the $n$-th time step (see Fig.\ref{figgrid}). The distribution function update can be written as 
\begin{eqnarray}
f_{\alpha,i}^{n+1,j} & = & f_{\alpha,i}^{n,j}-\frac{c_{\alpha x}\delta_{t}}{\delta_{x}}\left[\mathcal{F}_{\alpha,i+1/2}^{n,j}-\mathcal{F}_{\alpha,i-1/2}^{n,j}\right]\\
 & - & \frac{c_{\alpha y}\delta_{t}}{\delta_{y}}\left[\mathcal{F}_{\alpha,i}^{n,j+1/2}-\mathcal{F}_{\alpha,i}^{n,j-1/2}\right]\nonumber \\
 & + & \frac{\delta_{t}}{\tau}(f_{\alpha,i}^{ES,n,j}-f_{\alpha,i}^{n,j})+g_{\alpha}\delta_{t},\nonumber 
\end{eqnarray}
where $\delta_{x}$ and $\delta_{y}$ are the grid spacing in the $x$ and $y$ directions, respectively, 
and $\delta_{t}$ is the time step; $c_{\alpha x}$ and $c_{\alpha y}$
denote the particle velocity components at the $x$ and $y$ coordinates.
The outgoing and incoming fluxes at the node $(i,j)$ (see Fig.\ref{figgrid}) are 
\begin{equation}
\mathcal{F}_{\alpha,i+1/2}^{n,j}=f_{\alpha,i}^{n,j}+\frac{1}{2}\left(1-\frac{c_{\alpha x}\delta_{t}}{\delta_{x}}\right)\left[f_{\alpha,i+1}^{n,j}-f_{\alpha,i}^{n,j}\right]\Psi\left(\Theta_{\alpha,i}^{n}\right),
\end{equation}
\begin{equation}
\mathcal{F}_{\alpha,i-1/2}^{n,j}=\mathcal{F}_{\alpha,(i-1)+1/2}^{n,j},
\end{equation}
\begin{equation}
\mathcal{F}_{\alpha,i}^{n,j+1/2}=f_{\alpha,i}^{n,j}+\frac{1}{2}\left(1-\frac{c_{\alpha y}\delta_{t}}{\delta_{y}}\right)\left[f_{\alpha,i}^{n,j+1}-f_{\alpha,i}^{n,j}\right]\Psi\left(\Theta_{\alpha}^{n,j}\right),
\end{equation}
\begin{equation}
\mathcal{F}_{\alpha,i}^{n,j-1/2}=\mathcal{F}_{\alpha,i}^{n,(j-1)+1/2},
\end{equation}
where
\begin{equation}
\Theta_{\alpha,i}^{n}=\frac{f_{\alpha,i}^{n,j}-f_{\alpha,i-1}^{n,j}}{f_{\alpha,i+1}^{n,j}-f_{\alpha,i}^{n,j}},
\end{equation}
\begin{equation}
\Theta_{\alpha}^{n,j}=\frac{f_{\alpha,i}^{n,j}-f_{\alpha,i}^{n,j-1}}{f_{\alpha,i}^{n,j+1}-f_{\alpha,i}^{n,j}},
\end{equation}
and the minmod flux limiter is given by
\begin{equation}
\Psi\left(\Theta\right)=\max\left[0,\min(1,\Theta)\right].
\end{equation}
For simplicity, only the formulae for $c_{\alpha y}>0$ and $c_{\alpha x}>0$ are given. 

\begin{figure}
\begin{centering}
\includegraphics[width=0.48 \textwidth]{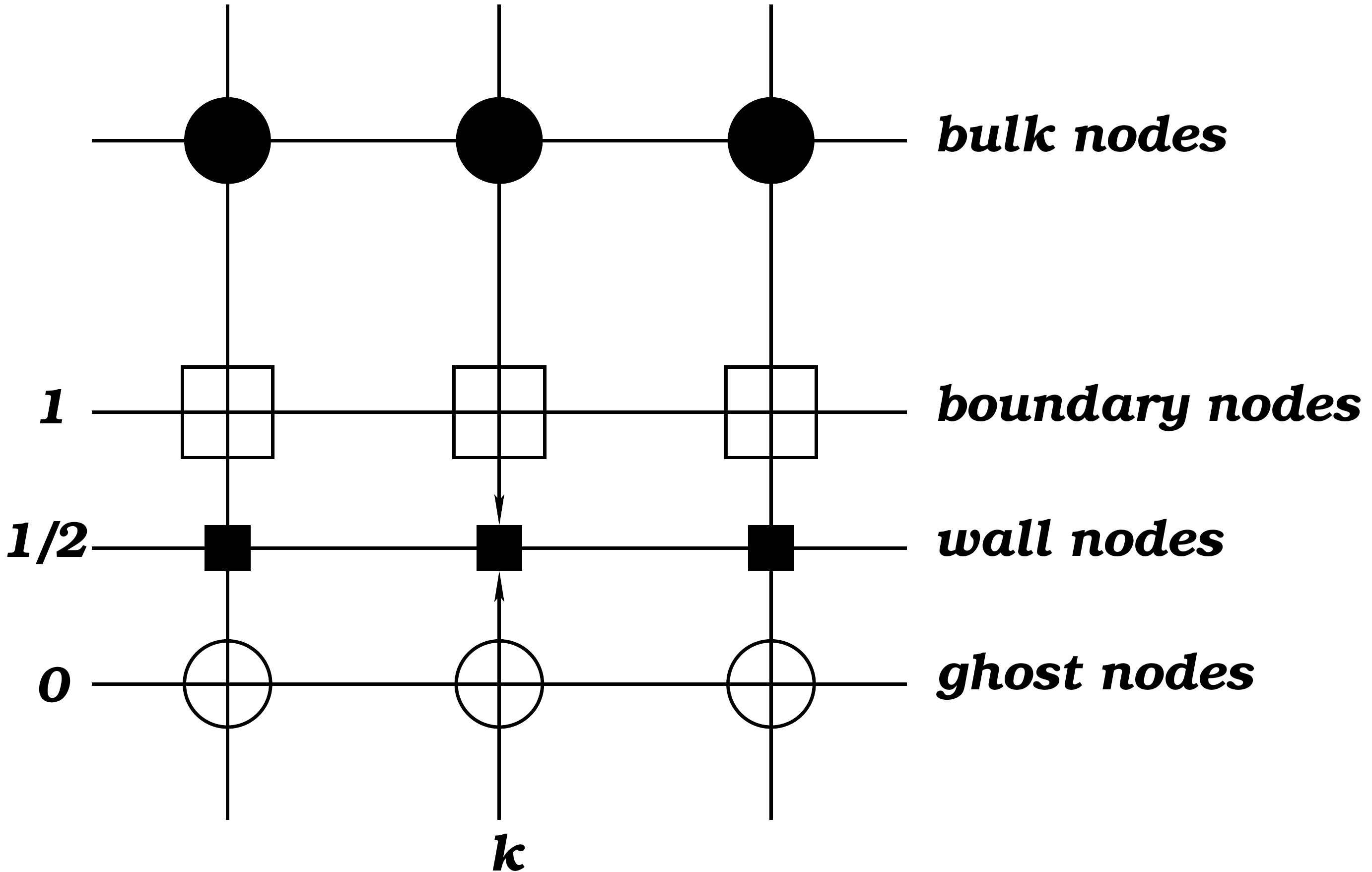} \caption{Schematic illustration of wall boundary treatment.\label{figboundary}}
\end{centering} 
\end{figure}
\subsection{Boundary condition}
Gas-wall interactions are captured by boundary conditions. 
The most popular boundary condition is the Maxwell model 
\citep{Cercignani2000,1879RSPT..170..231C}, 
in which a fraction ($1-\gamma$) of gas particles are assumed to undergo
specular reflection while the remaining particles are
diffusely reflected. In the limit of $\gamma=1$, 
the reflection becomes fully diffuse.

As lattice models are derived from continuous kinetic equations, their
boundary conditions may be obtained from the
continuous kinetic theory \citep{PhysRevE.66.026311}. It has already 
been shown that lattice models with diffuse type wall
boundary can describe a range of rarefied
effects~\citep[see, e.g.,][]{Meng2011b, Kim20088655,yudistiawan:016705}. In this
work, our primary interest is the model accuracy; as a result 
we will only implement the Maxwellian diffuse wall boundary.
   
The boundary condition employed in this work is  Version 1 
of boundary conditions in~\citet{Sofonea2009}, which is briefly described below.
Firstly, the truncated Maxwellian distribution function can be written as
\begin{eqnarray}
\label{feq0}
f^{0}_{\alpha} & = & \rho S(T,\bm{u)} \nonumber\\ 
 & = & \rho w_{\alpha}\left\{ 1+c_{i}u_{i}+\frac{1}{2}\left[(c_{i}u_{i})^{2}-u_{i}u_{i}+(T-1)(c_{i}c_{i}-D)\right]\right.\\
 & + & \frac{c_{i}u_{i}}{6}[(c_{i}u_{i})^{2}-3u_{i}u_{i}+3(T-1)(c_{i}c_{i}-D-2)]\nonumber\\
 & + & \frac{1}{24}[(c_{i}u_{i})^{4}-6(u_{i}c_{i})^{2}u_{j}u_{j}+3(u_{j}u_{j})^{2}]\nonumber\\
 & + &
 \frac{T-1}{4}[(c_{i}c_{i}-D-2)((u_{i}c_{i})^{2}-u_{i}u_{i})-2(u_{i}c_{i})^{2}]
 \nonumber\\
 & + & \frac{(T-1)^{2}}{8}\left.\left[(c_{i}c_{i})^{2}-2(D+2)c_{i}c_{i}+D(D+2)\right]\right\}.\nonumber 
\end{eqnarray}
As the discretization is conducted along a Cartesian coordinate system
(see Fig.\ref{figboundary}), the wall boundary condition can be written  as 
\begin{equation}
f_{\alpha,k}^{0}=\rho_{W,k}S(T_{W,k},\bm{u}_{W,k})\mbox{\ \ \ }\bm{\xi}_{\alpha}\cdot\bm{n}>0,
\end{equation}
where
\begin{equation}
{\displaystyle \rho_{W,k}=\frac{\sum\limits _{(\bm{\xi}_{\alpha}\cdot\bm{n})<0}\left|\bm{\xi}_{\alpha}\cdot\bm{n}\right|f_{\alpha,k}^{1}}{\sum\limits _{(\bm{\xi}_{\alpha}\cdot\bm{n})>0}\left|\bm{\xi}_{\alpha}\cdot\bm{n}\right|S(T_{W,k},\bm{u}_{W,k})},}
\end{equation}
Here,  subscript $W$ denotes the computational nodes at the wall, 
$\rho_{W,k}$ is the density on the wall nodes $k$ (see Fig.\ref{figboundary}),
$T_{W,k}$ is the temperature, $\bm{u}_{W,k}$ the velocity and $\bm{n}$
the inward unit vector normal to the wall. 
The distribution function
at the ghost nodes is assumed to be identical to those on the corresponding wall nodes.

\section{Validation}
We have validated the proposed scheme using a variety of flows, both in the low-Mach and high-Mach number limit. 
Unless otherwise stated, our results have been obtained using a 11th-order of discrete velocity set, which represents a 
good compromise between accuracy (expected to be accurate at
the Burnett level, see Sec.\ref{remarkaccuracy}) and computational efficiency. Our numerical results will be compared with DSMC 
and LVDSMC~\citep[see][]{Homolle2007,PhysRevE.79.056711,pof2011} simulations, 
while in some cases, numerical solutions of the ES-BGK equation and analytical results of the R13 model will also be presented. 

The flows considered here are one-dimensional and are confined by two infinite parallel plates in the $x-z$ plane. 
The distance between the two plates (in the $y$ direction) 
is $L$ on which the Knudsen number
\begin{equation}
K_D=\sqrt{\frac{\pi}{2}} Kn 
\end{equation}
is based.
Comparisons with the hard-sphere gas will be performed by setting the Prandtl
number to $2/3$, i.e. $b=-1/2$. For the BGK gas, the Prandtl
number is unity which corresponds to
$b=0$. Furthermore, the gas has the heat capacity
  $C_p=5/2$, i.e., the spatial dimension $D$ is set to be
  $3$, see Eqs.~(\ref{eq:mac}) and (\ref{feq0}). Our simulation results ($D=3$) are described below. 
\subsection{Low Mach number flows}
\label{lowMa}

\begin{figure*}
\begin{center}
\includegraphics[width=0.48 \textwidth,height=0.36 \textwidth]{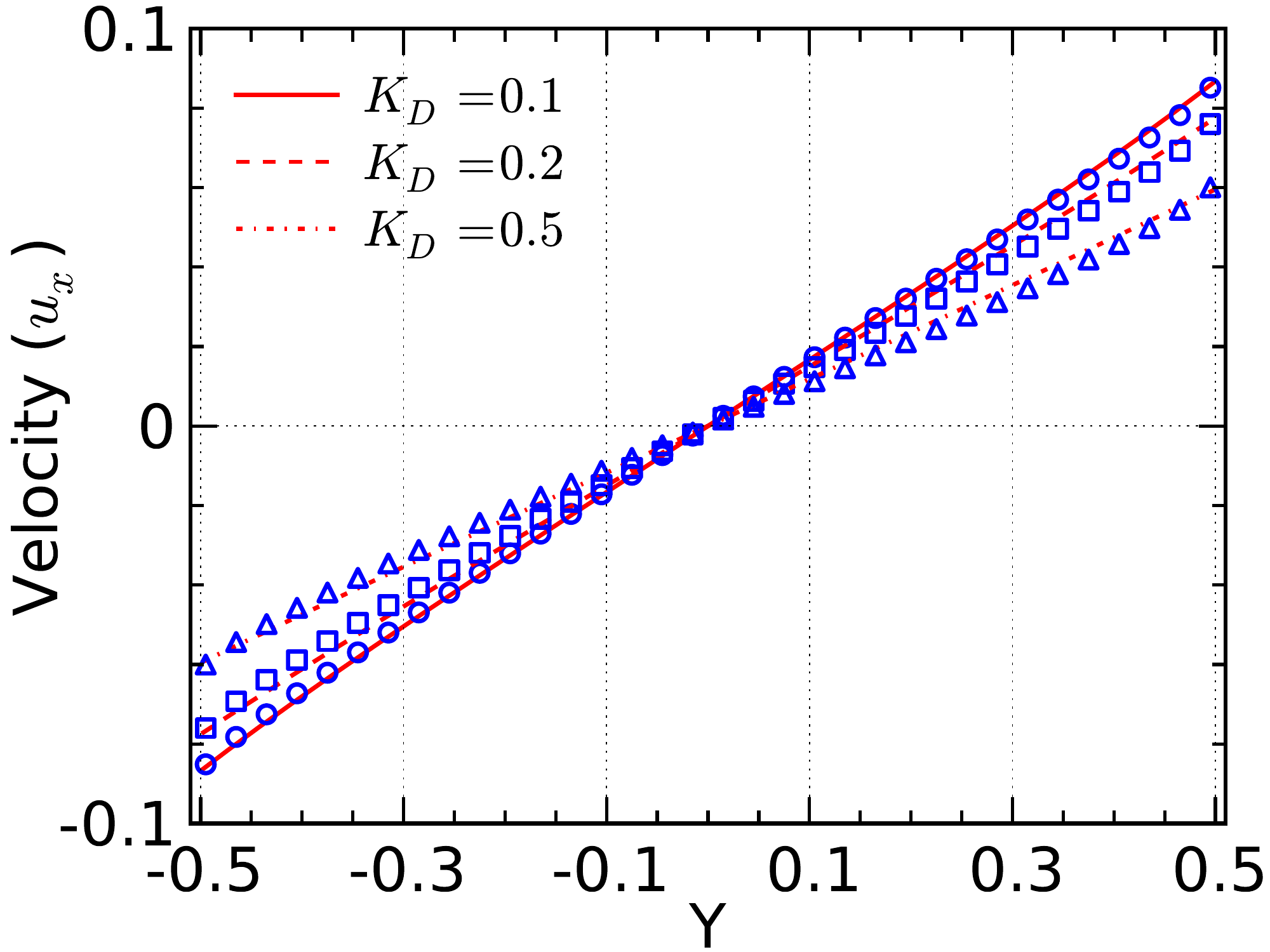}  
\includegraphics[width=0.48 \textwidth,height=0.36
\textwidth]{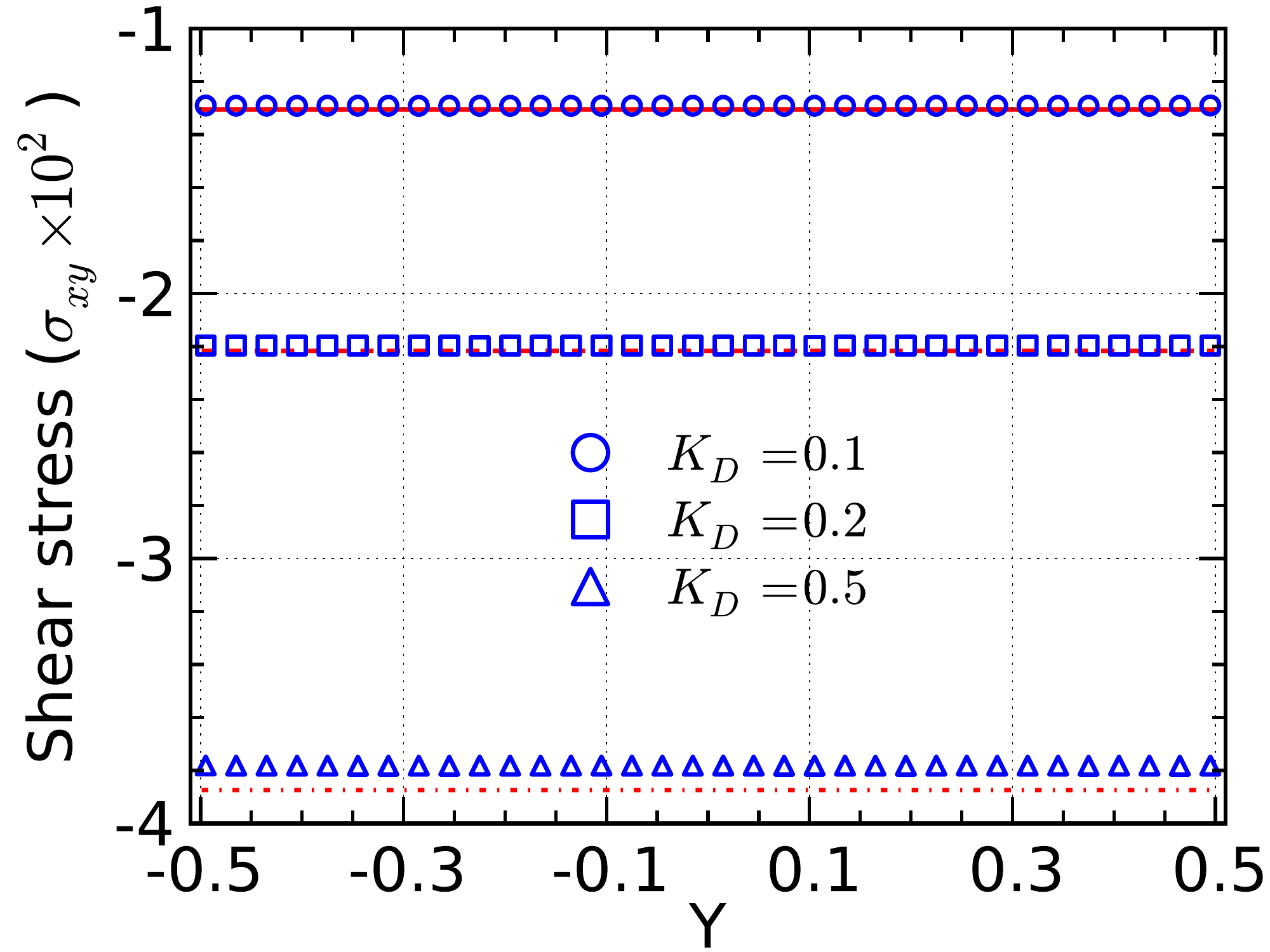}
\caption{The velocity and shear stress profiles  for Couette
flow of a hard-sphere gas at $K_D=0.1$, $0.2$ and $0.5$. The symbols denote DSMC data
  and the lines represent the lattice ES-BGK results. The wall velocities are $u_w=\pm 0.1$. \label{couetteutlowma}} 
\end{center}
\end{figure*}

\begin{figure*}
\begin{center}
\includegraphics[width=0.48 \textwidth,height=0.36 \textwidth]{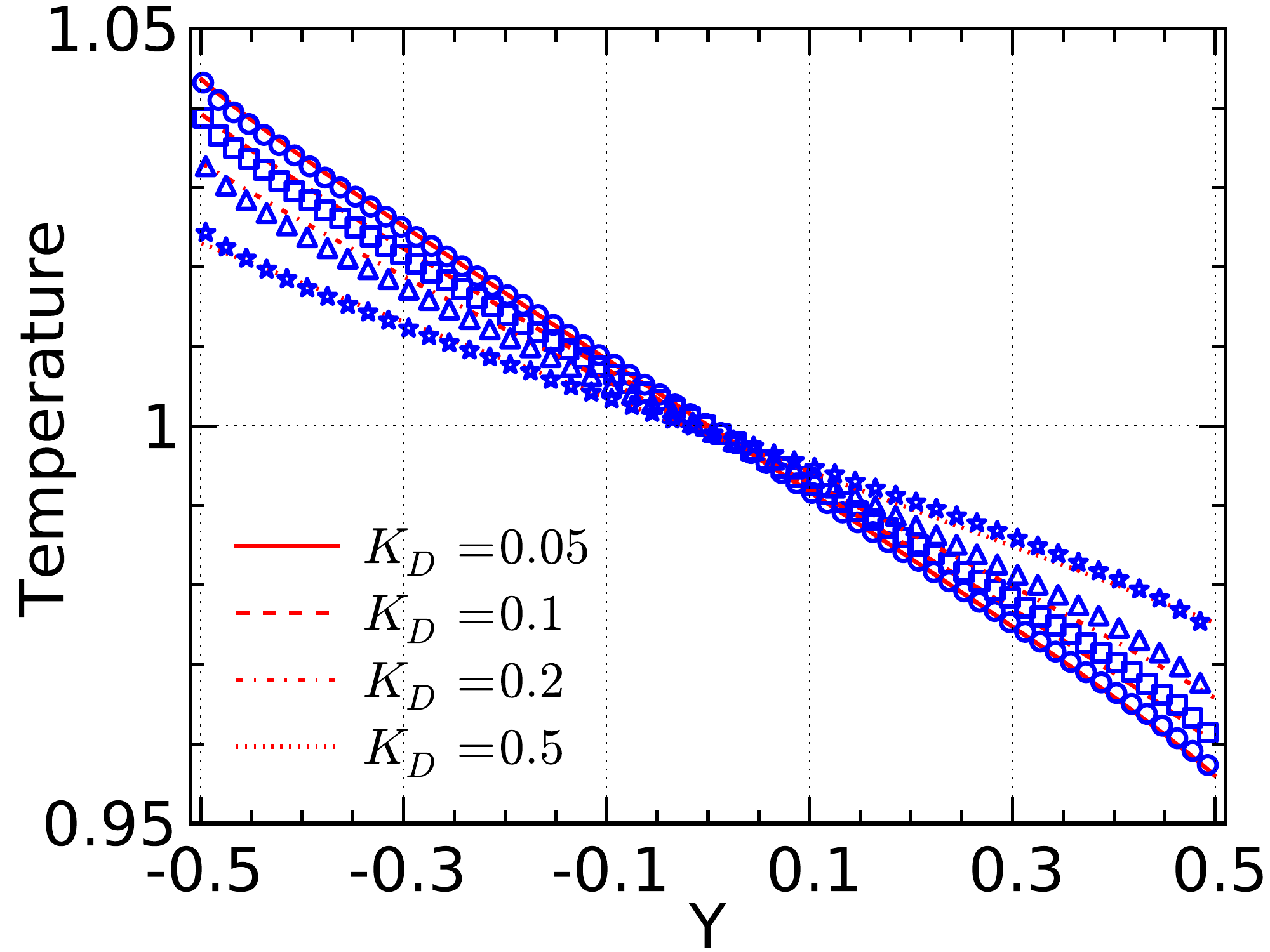}  
\includegraphics[width=0.48 \textwidth,height=0.36
\textwidth]{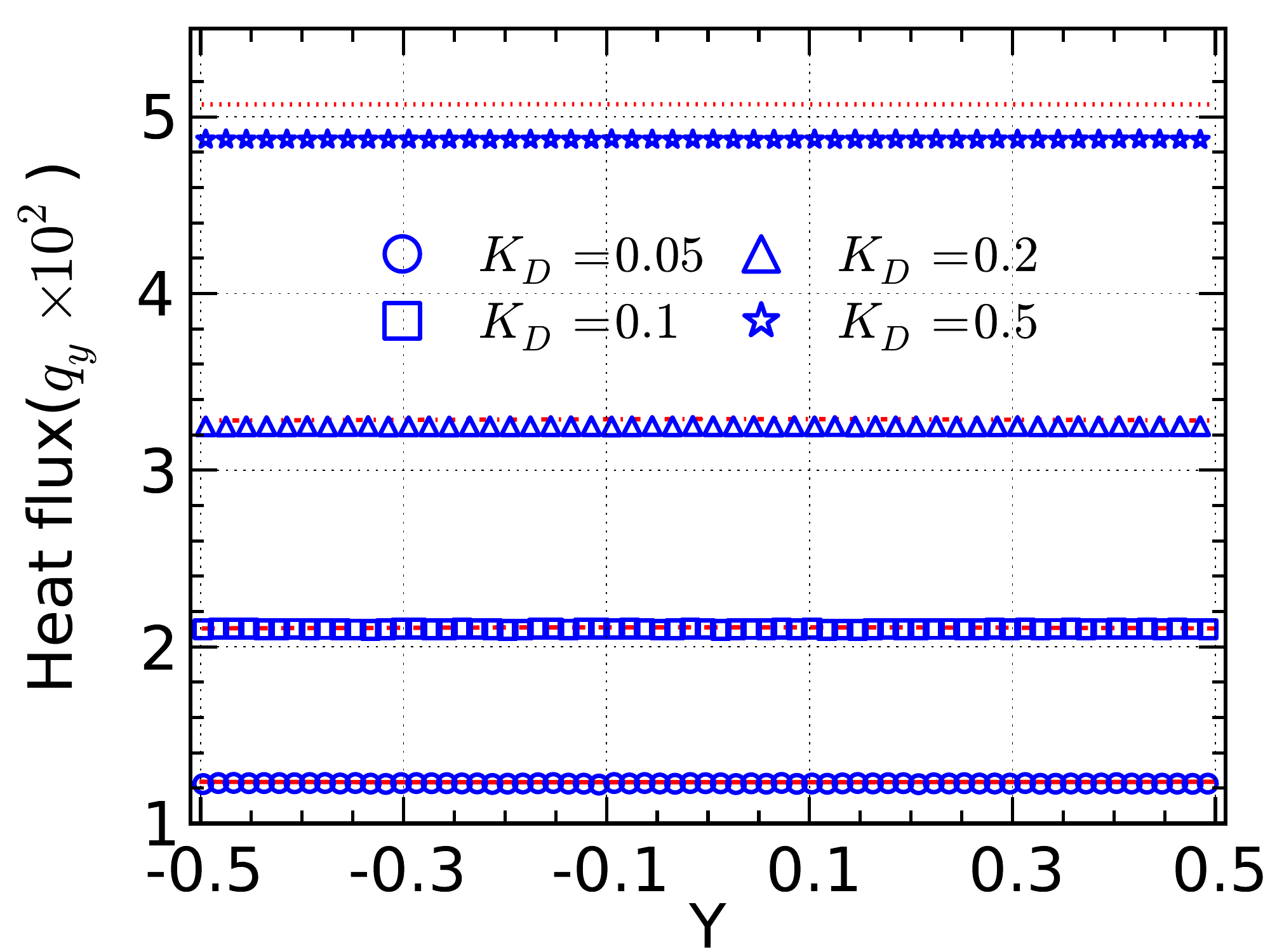}
\caption{The temperature and heat flux profiles for a Fourier problem 
for a hard-sphere gas at $K_D=0.05, 0.1$, $0.2$ and $0.5$
with the wall temperature difference $0.1$. The symbols denote DSMC data
  and the lines represent the lattice ES-BGK results. \label{linearfouier}} 
\end{center}
\end{figure*}

Flows in the low-Mach number limit are of particular interest since they are most often encountered  in micro/nanofluidic 
applications ~\citep{cerc2006,Hadjiconstantinou2006}.  Here we provide validation of our methodology using traditional  
Couette and Fourier flows. 
We also present solutions of an  unsteady boundary heating problem that features coupled momentum and heat transfer 
and thus requires an accurate representation of the ratio of the associated transport coefficients, namely the Prandtl number.
 This problem also allows us to investigate the limitations of the proposed model in the presence of kinetic effects due to 
flow unsteadiness.  
\subsubsection{Steady Couette and Fourier Flows}
Figures \ref{couetteutlowma} and \ref{linearfouier} show comparison between our lattice results and DSMC simulations for a 
Couette and 
a Fourier heat tranfer problem respectively. In the Couette flows, the walls move with velocities $u_w=\pm 0.1$, while 
in the Fourier problems the wall temperatures are $T_w=1\pm 0.05$.
Fig.~\ref{linearfouier} shows that the proposed lattice ES-BGK model 
can be used to approximate the hard-sphere gas thermal conductivity by setting the Prandtl number to 
$2/3$.
Both figures show that the agreement in the slip-flow regime is excellent; as $K_D$ increases into the 
transition regime, a small amout of error is evident. This error increases further 
as $K_D$ increases, but remains very reasonable for $K_D<0.5$; for example,  for both flows, at 
$K_D=0.5$ the error is less than 5\%.

\subsection{Unsteady boundary heating problems}
In this section we present results from an unsteady boundary heating problem of the kind studied 
in ~\citep{manelajfm}; these studies are 
motivated by the common occurrence of applications involving time-varying boundary temperatures 
in micro-electro-mechanical devices. 
In this problem, the two walls confining the gas  are heated uniformly
via prescribed time dependent 
temperatures of the form $T_w=1+F(t)$. Here, $F(t)$ is taken to be
a sinusoidal function $\sin(\theta t )$ with an amplitude ($0.002$ in
simulations) that is sufficiently small to justify a linear assumption. 
As this is a time-varying problem, 
another non-dimensional parameter needs to  be introduced 
~\citep{Manela2008,Manela2010}, namely the Strouhal number, defined as
\begin{equation}
St=\theta=\frac{\hat{\theta} L}{\sqrt{RT_0}}.
\end{equation}
Kinetic effects become important as both 
the Strouhal and Knudsen numbers increase as shown in Fig.~6 in~\citet{Manela2010}. 

Here, simulations will
be conducted for a range of Strouhal and Knudsen numbers. 
Prandtl numbers corresponding both to the hard-sphere and BGK model 
will be considered. The accuracy of the lattice results will be evaluated 
by comparison with simulation results obtained using the recently 
developed low-variance deviational simulation Monte Carlo (LVDSMC) method 
~\citep{Homolle2007,PhysRevE.79.056711,pof2011} because simulation of 
this low-Mach number flow by DSMC is prohibitively expensive. The LVDSMC method 
was developed to specifically address this DSMC limitation, namely the 
prohibitive computational cost associated with low-Ma and more generally low-signal 
flows. LVDSMC provides efficient numerical solutions 
of the Boltzmann equation for all Knudsen numbers ~\citep{wagner2008} 
and {\it arbitrarily small signals} by simulating only the deviation 
from equilibrium. This control-variate variance-reduction formulation introduces 
deterministic knowledge of a nearby equilibrium state, and thus 
significantly reduces the stastistical uncertainty 
associated with the Monte Carlo approach {\it without introducing any approximation}. The resulting 
computational benefits in the limit of low speed flows such as the one considered here are typically very large 
~\citep{Homolle2007}.

Following ~\citet{Manela2010}, comparison with LVDSMC results will be 
performed using the following two Knudsen numbers:
\begin{equation}
K_E= \frac{16}{5\sqrt{2\pi}} Kn,
\end{equation}
for the hard-sphere gas with $Pr=2/3$, and
\begin{equation}
K_B=\sqrt{\frac{8}{\pi}} Kn
\end{equation}
for the BGK gas with $Pr=1$. 
Details on the LVDSMC simulations of the problem studied here can be found in~\citet{Manela2008,Manela2010}.

\begin{figure*}
\begin{center}
\includegraphics[width=0.48 \textwidth,height=0.36 \textwidth]{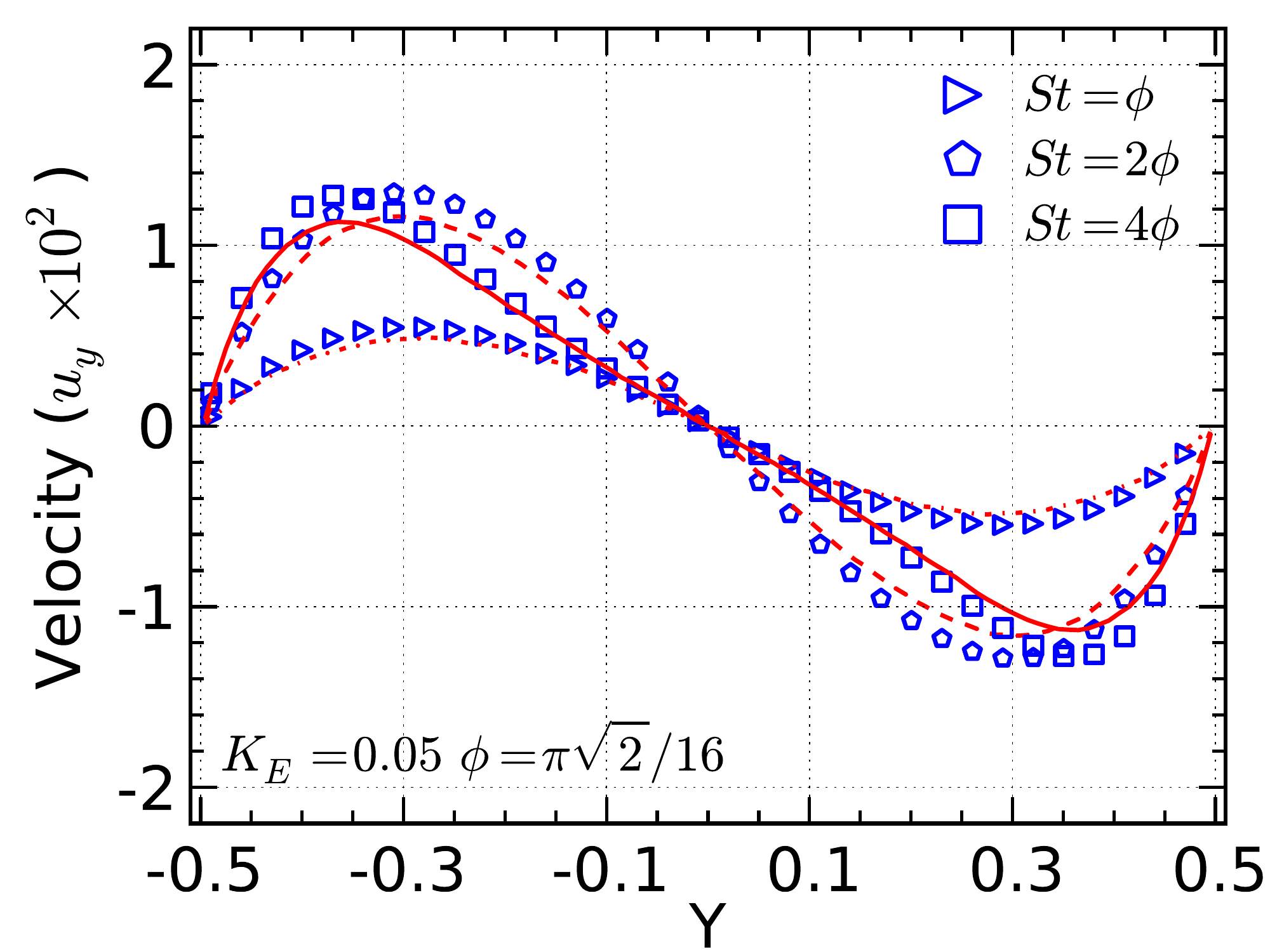}  
\includegraphics[width=0.48 \textwidth,height=0.36 \textwidth]{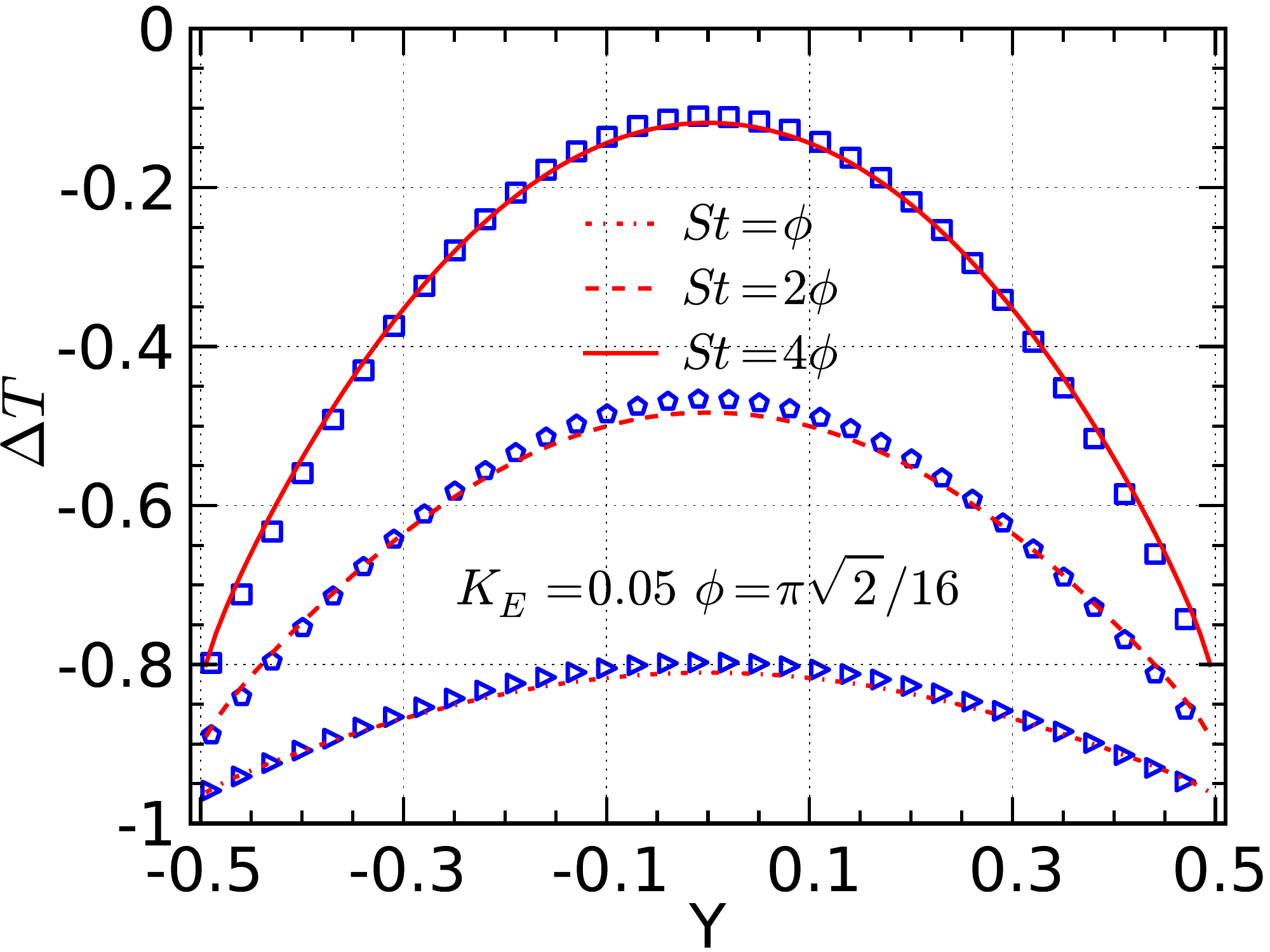}
\caption{The velocity and temperature perturbations for a
  hard-sphere gas subject to a sinusoidal heating at $t=3\pi/2$ at different Kn and St. The symbols correspond to the data of
  the lattice model ($Pr=2/3$) and the lines are the results of the LVDSMC method. \label{uhskn005}}
\end{center}
\end{figure*}

\begin{figure*}
\begin{center}
\includegraphics[width=0.48 \textwidth,height=0.36 \textwidth]{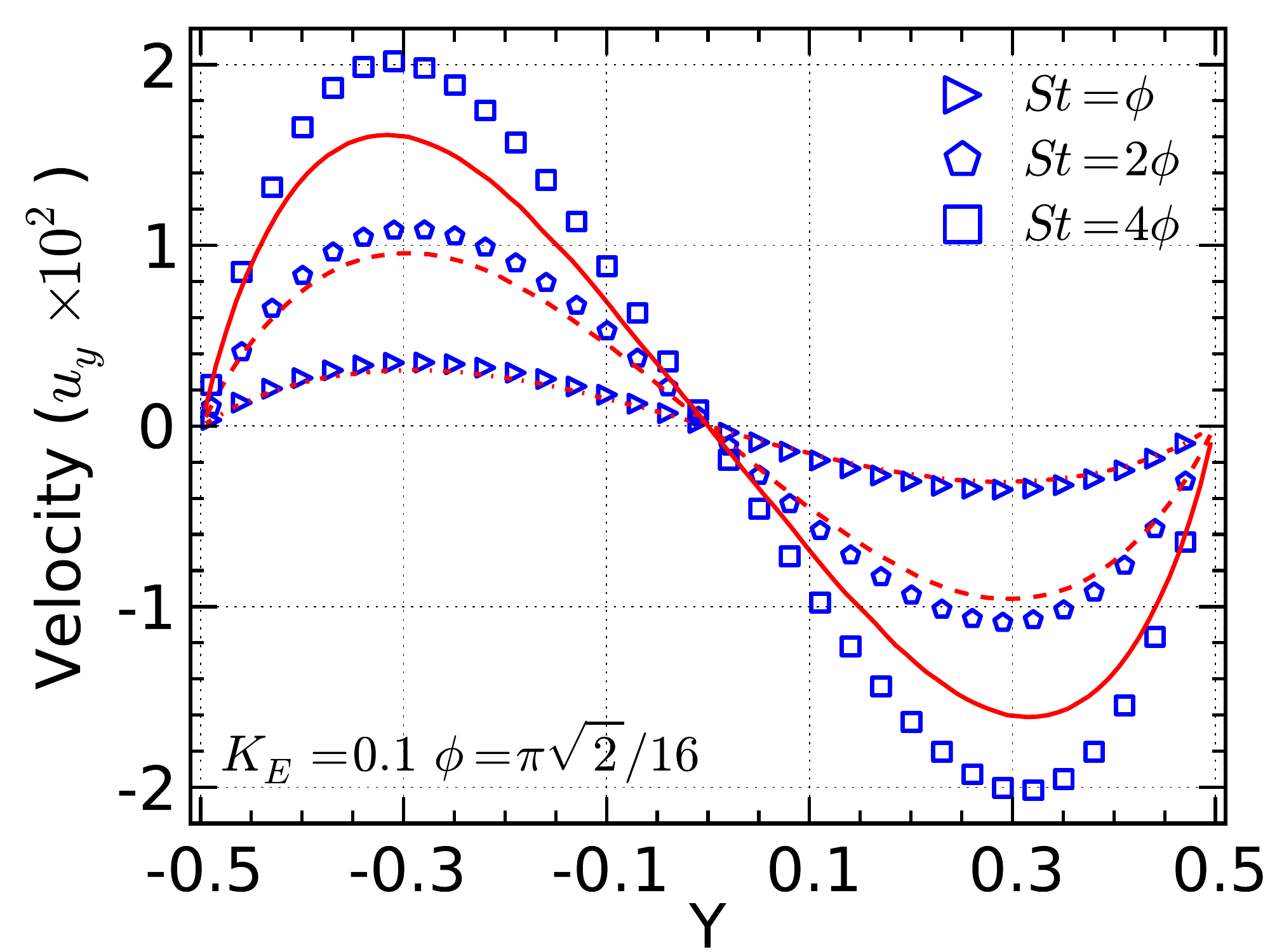}  
\includegraphics[width=0.48 \textwidth,height=0.36 \textwidth]{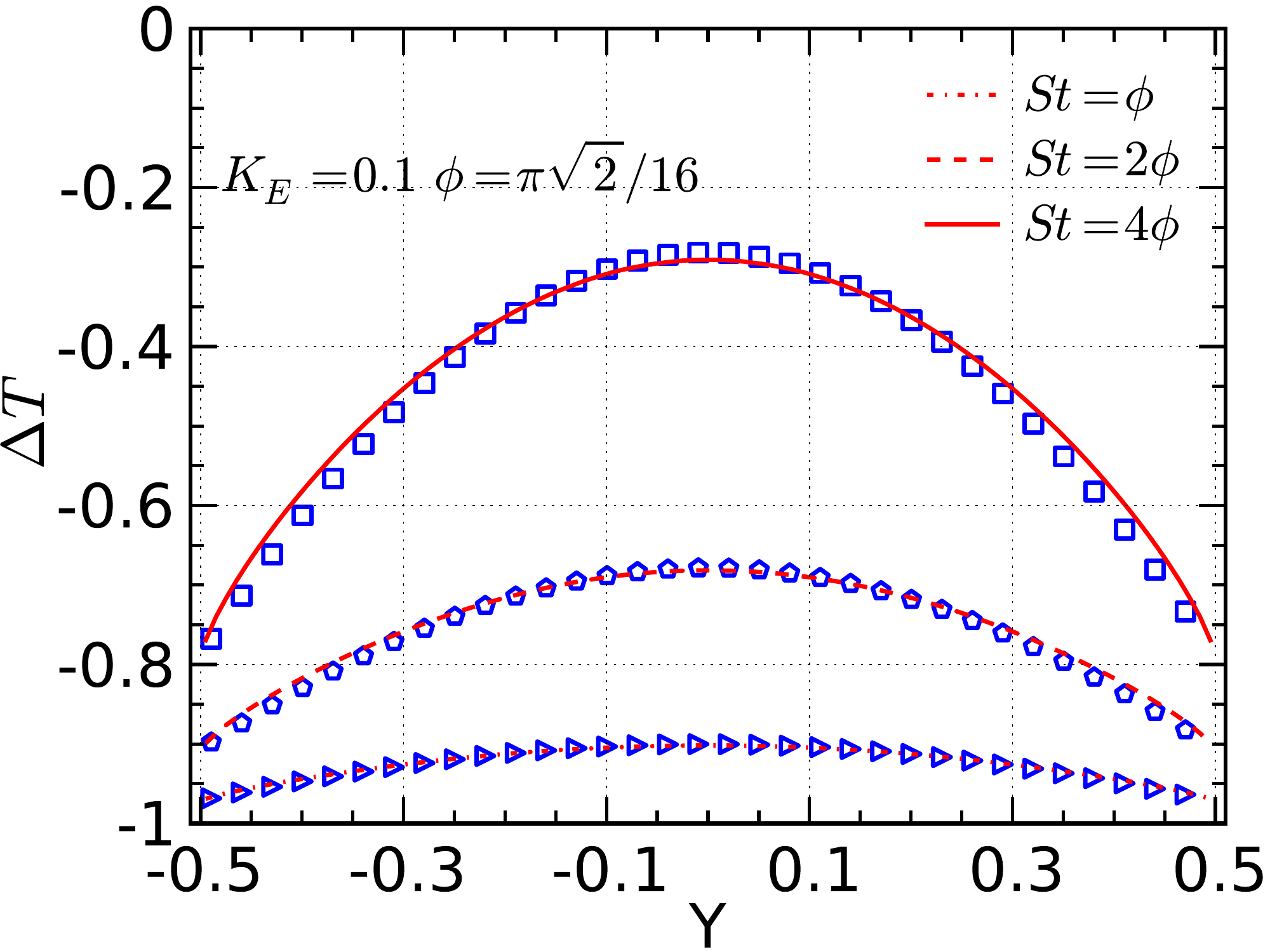}
\caption{The velocity and temperature perturbations ($\Delta T$) for a
  hardsphere gas subject to a sinusoidal heating at $t=3\pi/2$ at different Kn and St. The symbols correspond to the data of
  the lattice model ($Pr=2/3$) and the lines are the results of the LVDSMC method. \label{uhskn01} }
\end{center}
\end{figure*}

\begin{figure*}
\begin{center}
\includegraphics[width=0.48 \textwidth,height=0.36 \textwidth]{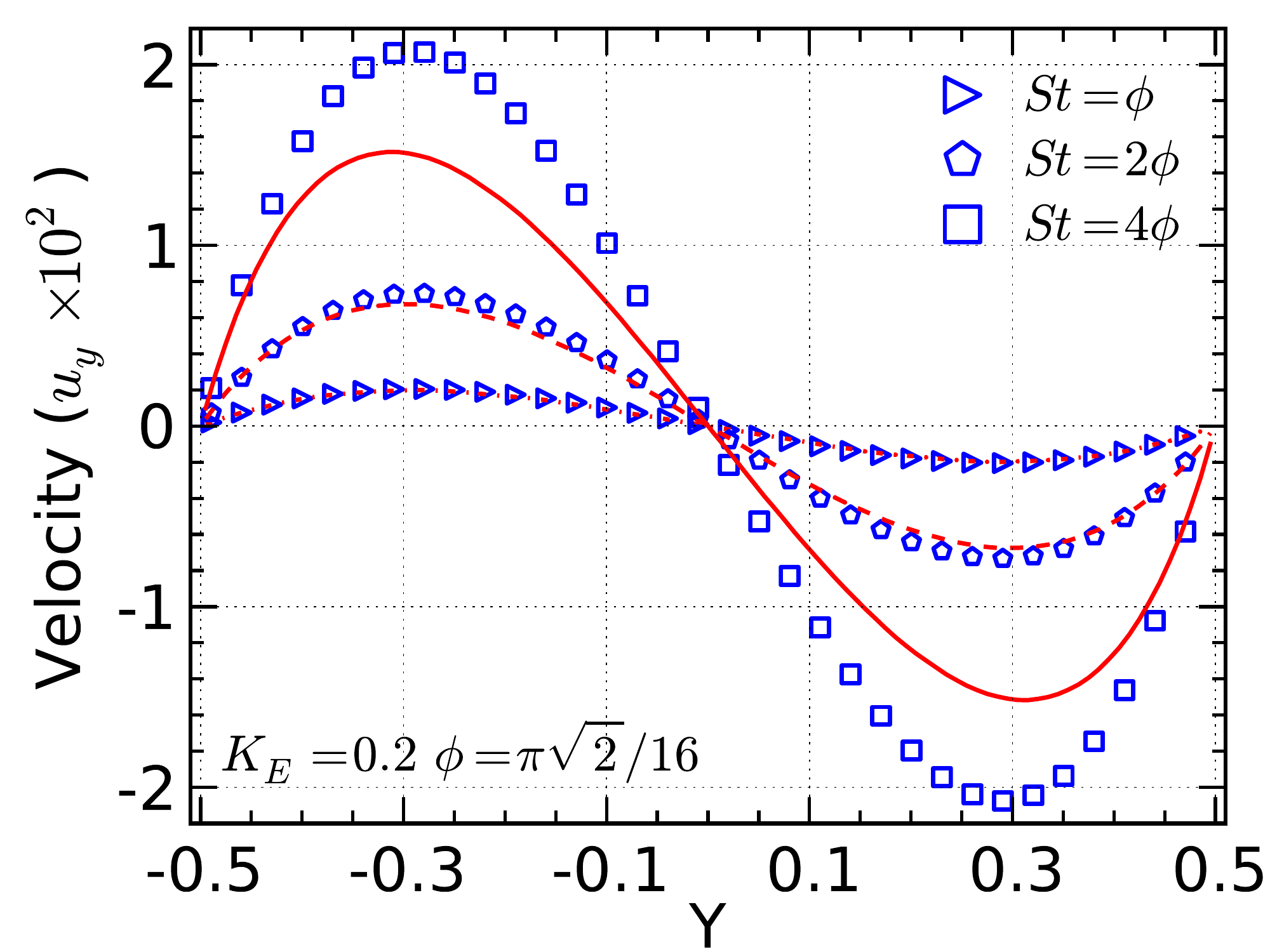}  
\includegraphics[width=0.48 \textwidth,height=0.36 \textwidth]{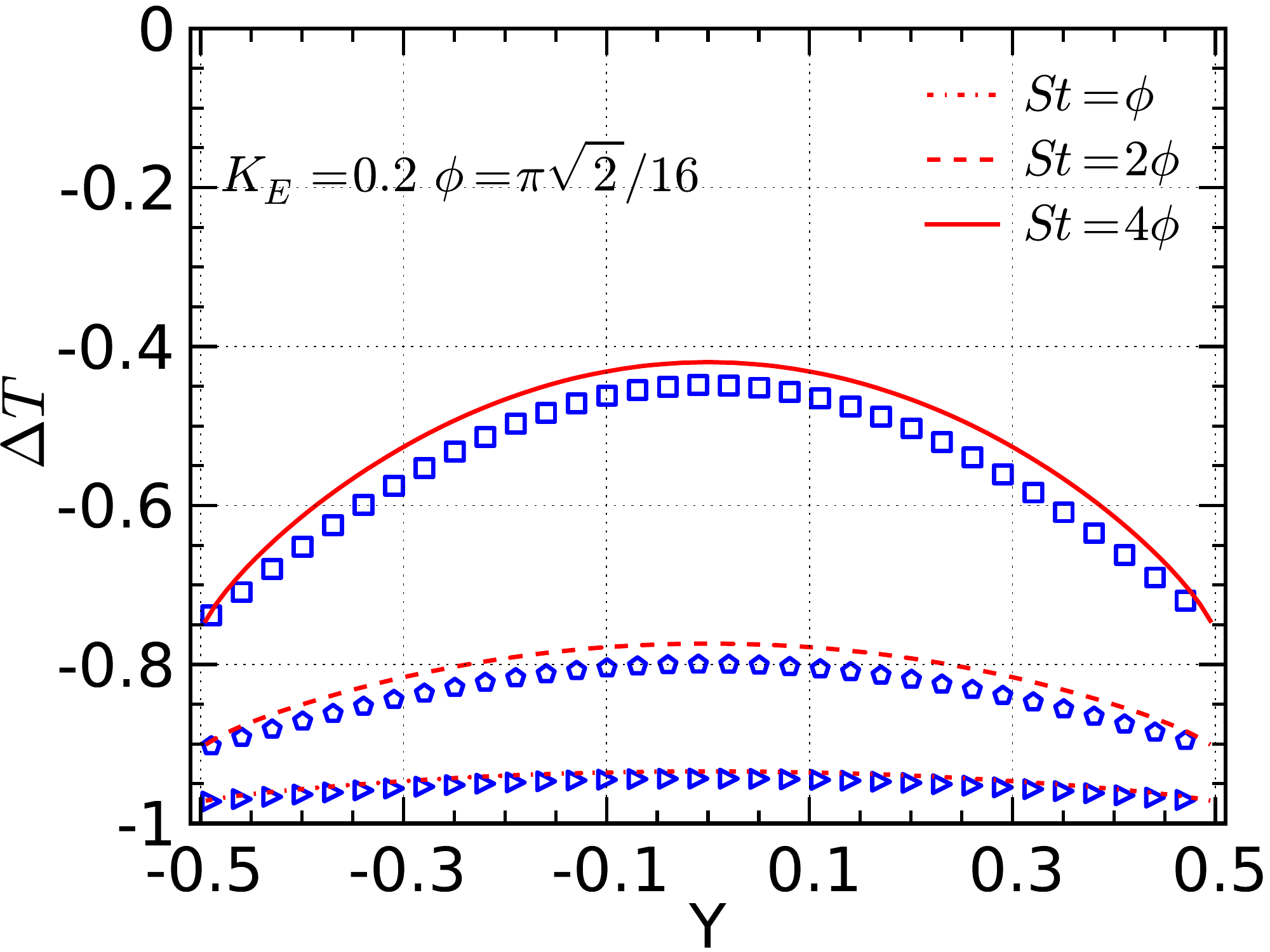}
\caption{The velocity and temperature perturbations for a
  hard-sphere gas subject to a sinusoidal heating at $t=3\pi/2$ at different Kn and St. The symbols correspond to the data of
  the lattice model ($Pr=2/3$) and the lines are the results of the LVDSMC method. \label{uhskn02} }
\end{center}
\end{figure*}

\begin{figure*}
\begin{center}
\includegraphics[width=0.48 \textwidth,height=0.36 \textwidth]{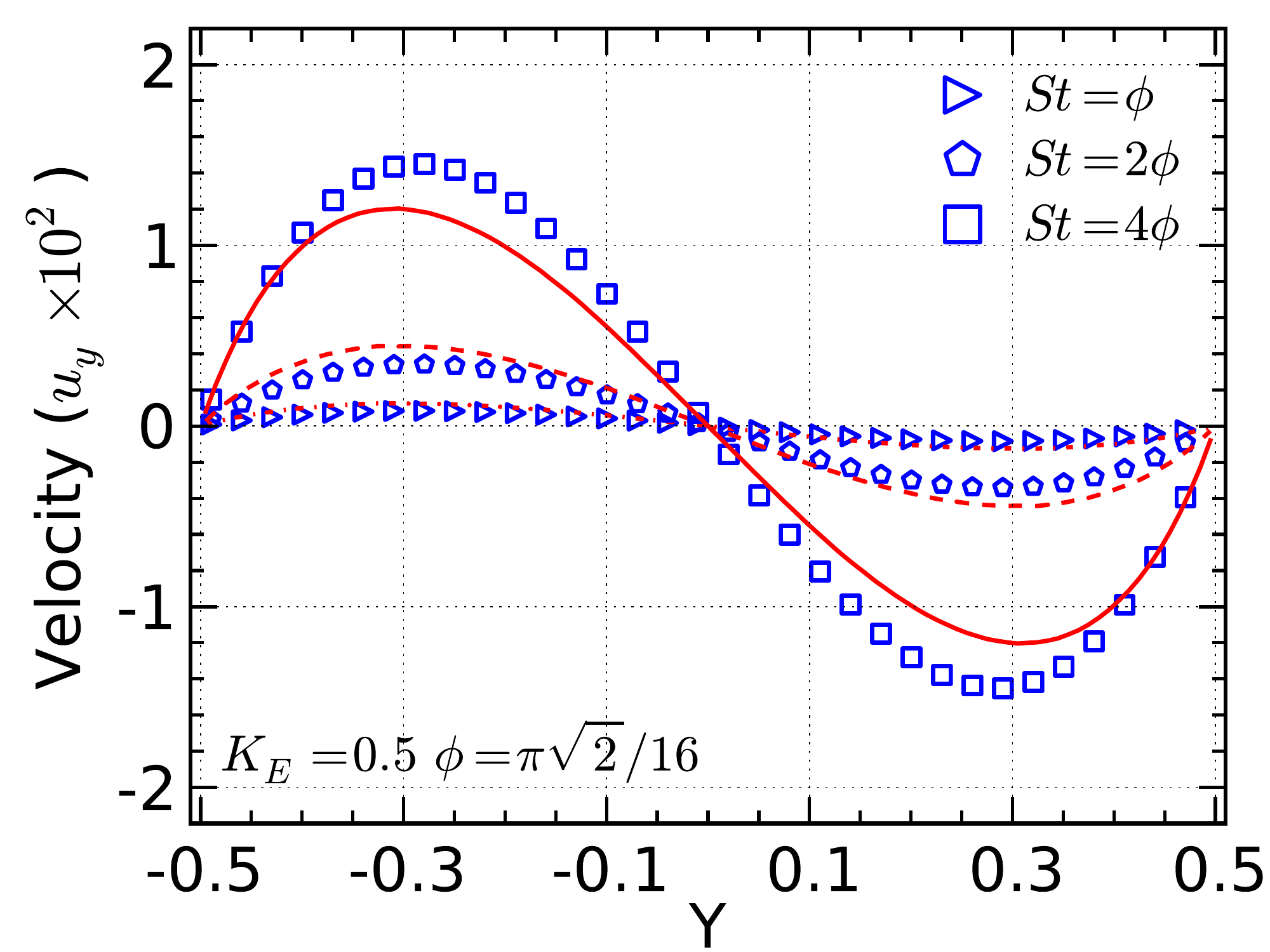}  
\includegraphics[width=0.48 \textwidth,height=0.36 \textwidth]{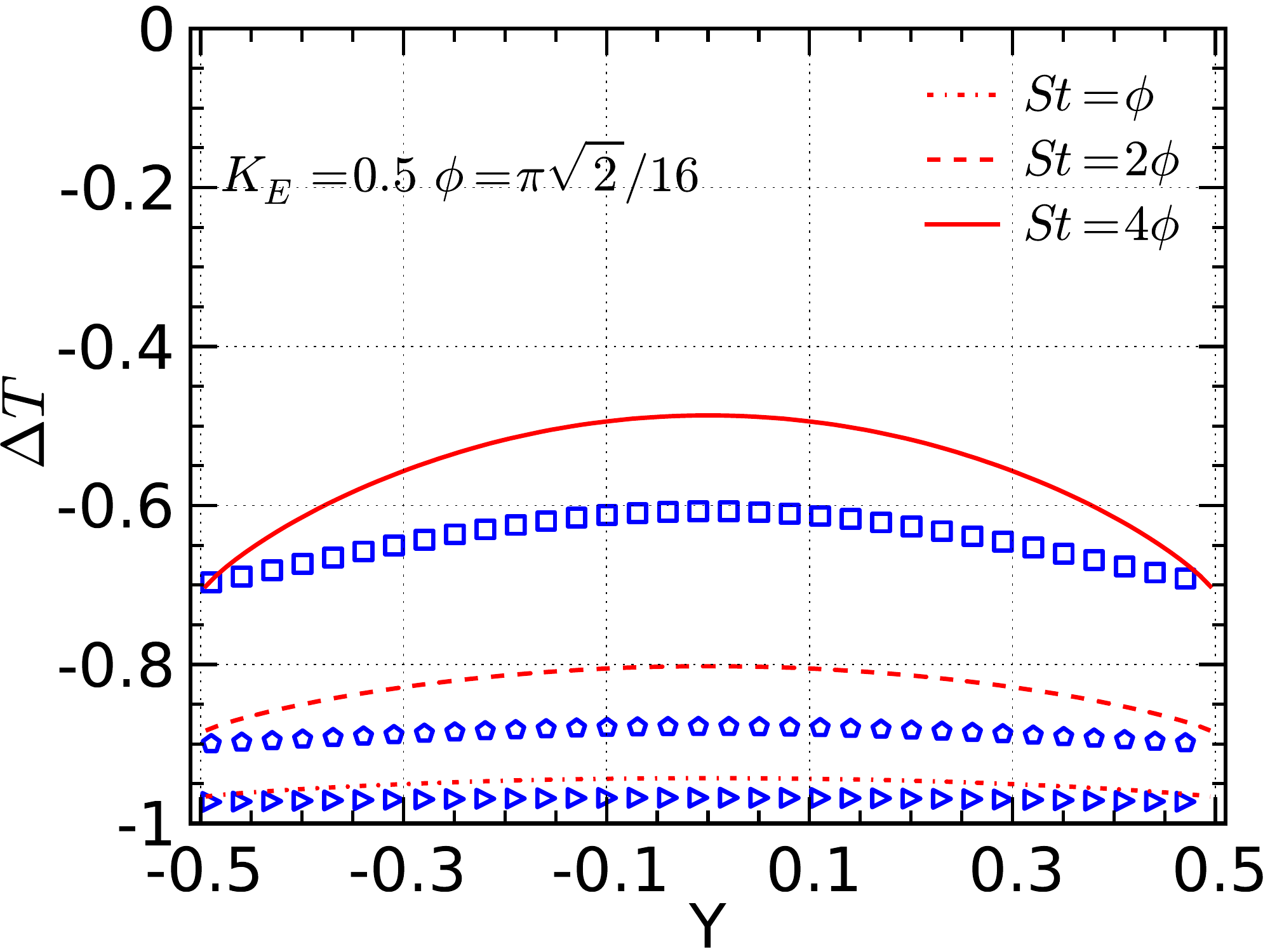}
\caption{The velocity and temperature perturbations  for a
  hard-sphere gas subject to a sinusoidal heating at $t=3\pi/2$ at different Kn and St. The symbols correspond to the data of
  the lattice model ($Pr=2/3$) and the lines are the results of the LVDSMC method. \label{uhskn05} }
\end{center}
\end{figure*}

\begin{figure*}
\begin{center}
\includegraphics[width=0.48 \textwidth,height=0.36 \textwidth]{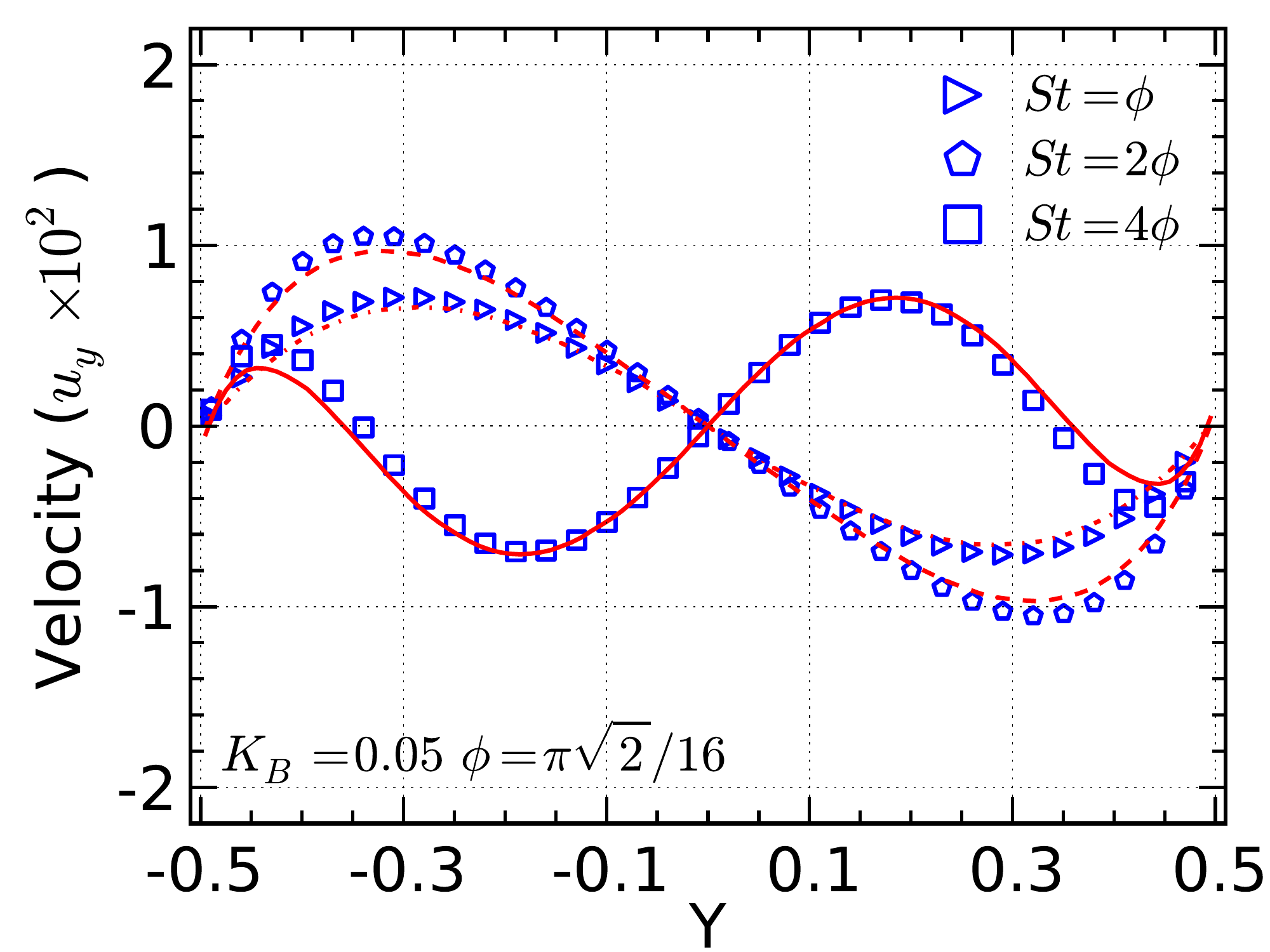}  
\includegraphics[width=0.48 \textwidth,height=0.36 \textwidth]{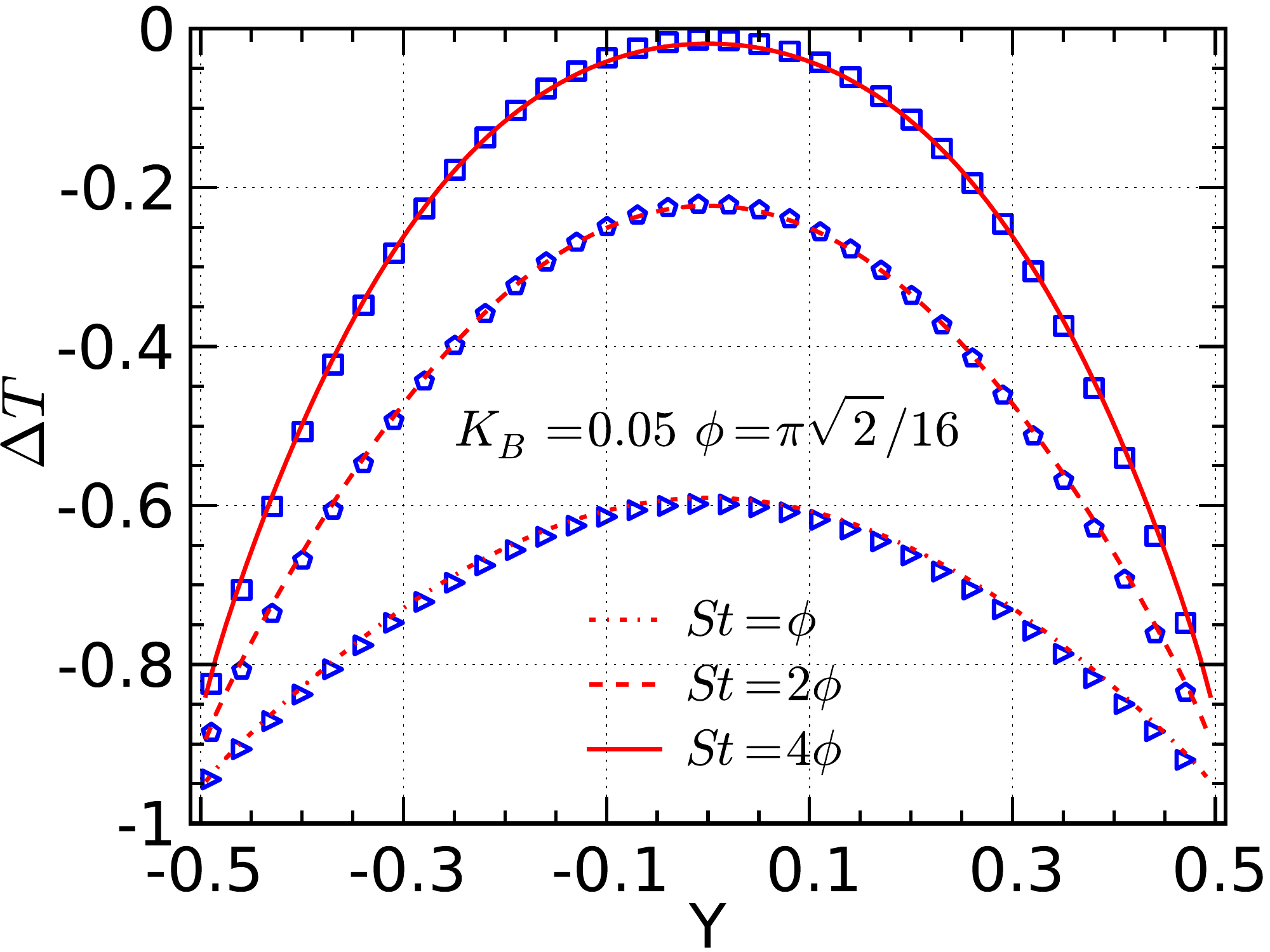}
\caption{The velocity and temperature perturbations for a BGK
  gas subject to a sinusoidal heating at $t=3\pi/2$ at different Kn and St. The symbols correspond to the data of the lattice
  model ($Pr=1$) and the lines are the results of the LVDSMC method.
\label{ubgkkn005}} 
\end{center}
\end{figure*}

\begin{figure*}
\begin{center}
\includegraphics[width=0.48 \textwidth,height=0.36 \textwidth]{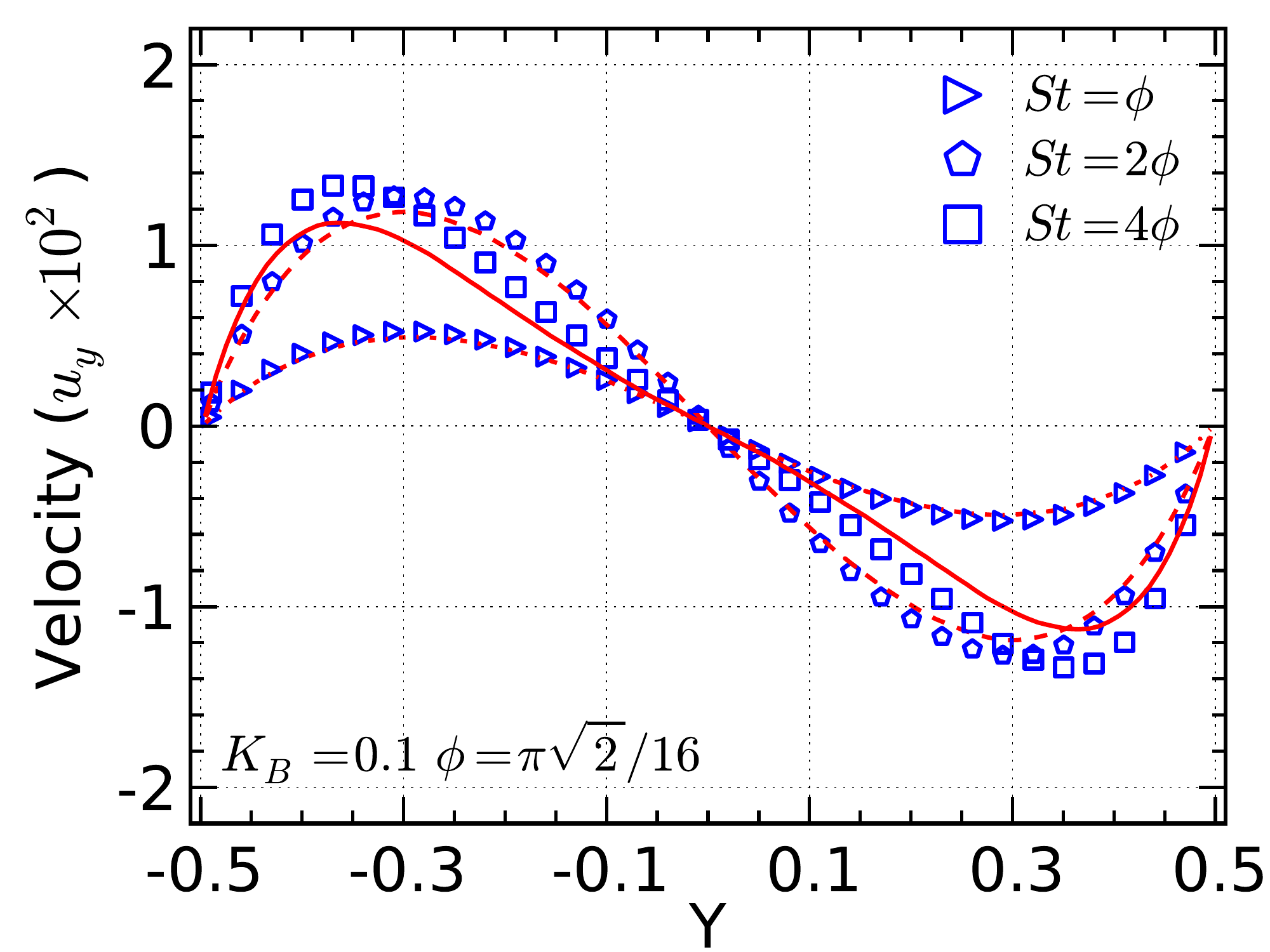}  
\includegraphics[width=0.48 \textwidth,height=0.36 \textwidth]{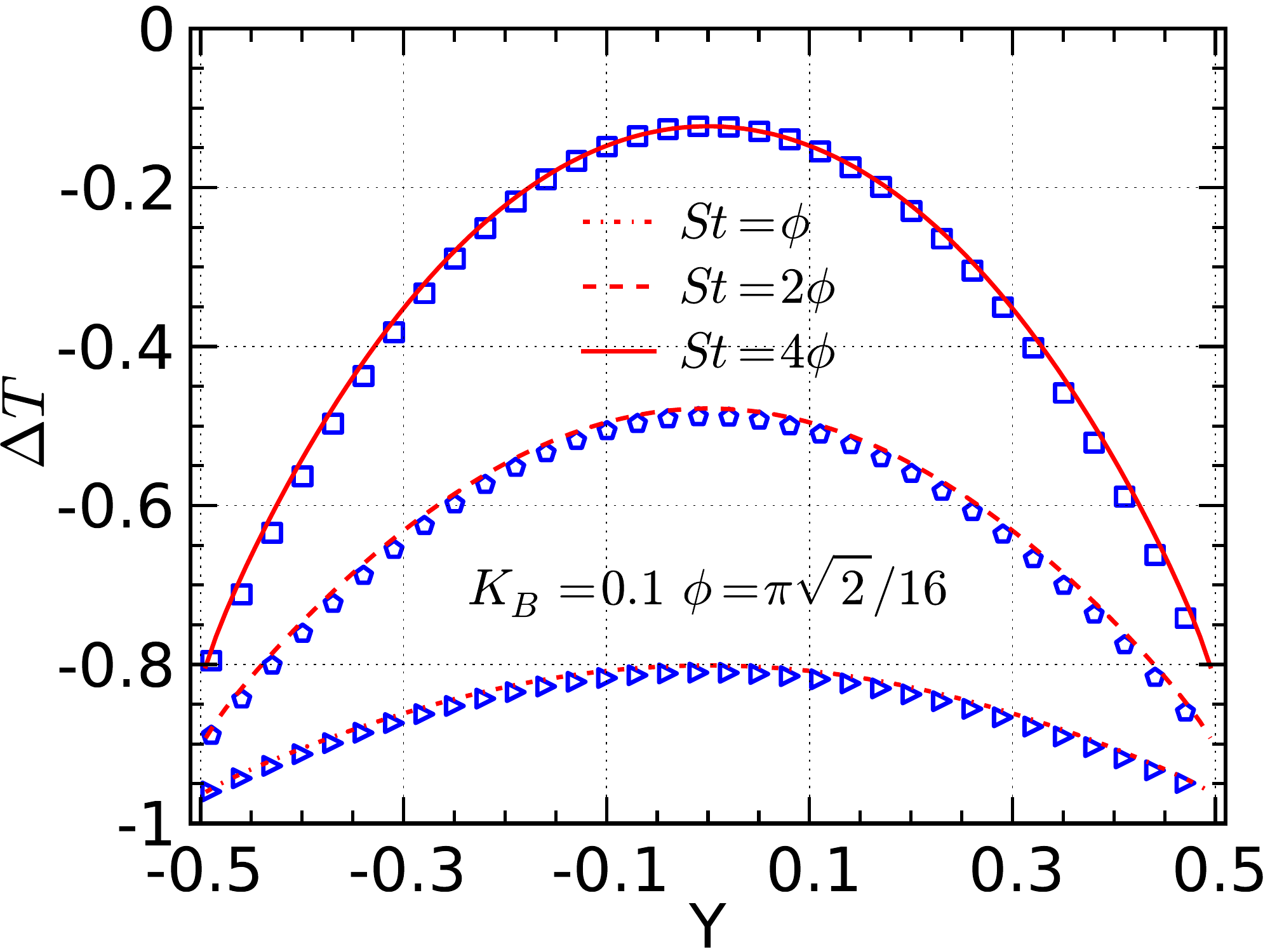}
\caption{The velocity and temperature perturbations ($\Delta T$) for a BGK
  gas subject to a sinusoidal heating at $t=3\pi/2$ at different Kn and St. The symbols correspond to the data of the lattice
  model ($Pr=1$) and the lines are the results of the LVDSMC method.
\label{ubgkkn01}} 
\end{center}
\end{figure*}

\begin{figure*}
\begin{center}
\includegraphics[width=0.48 \textwidth,height=0.36 \textwidth]{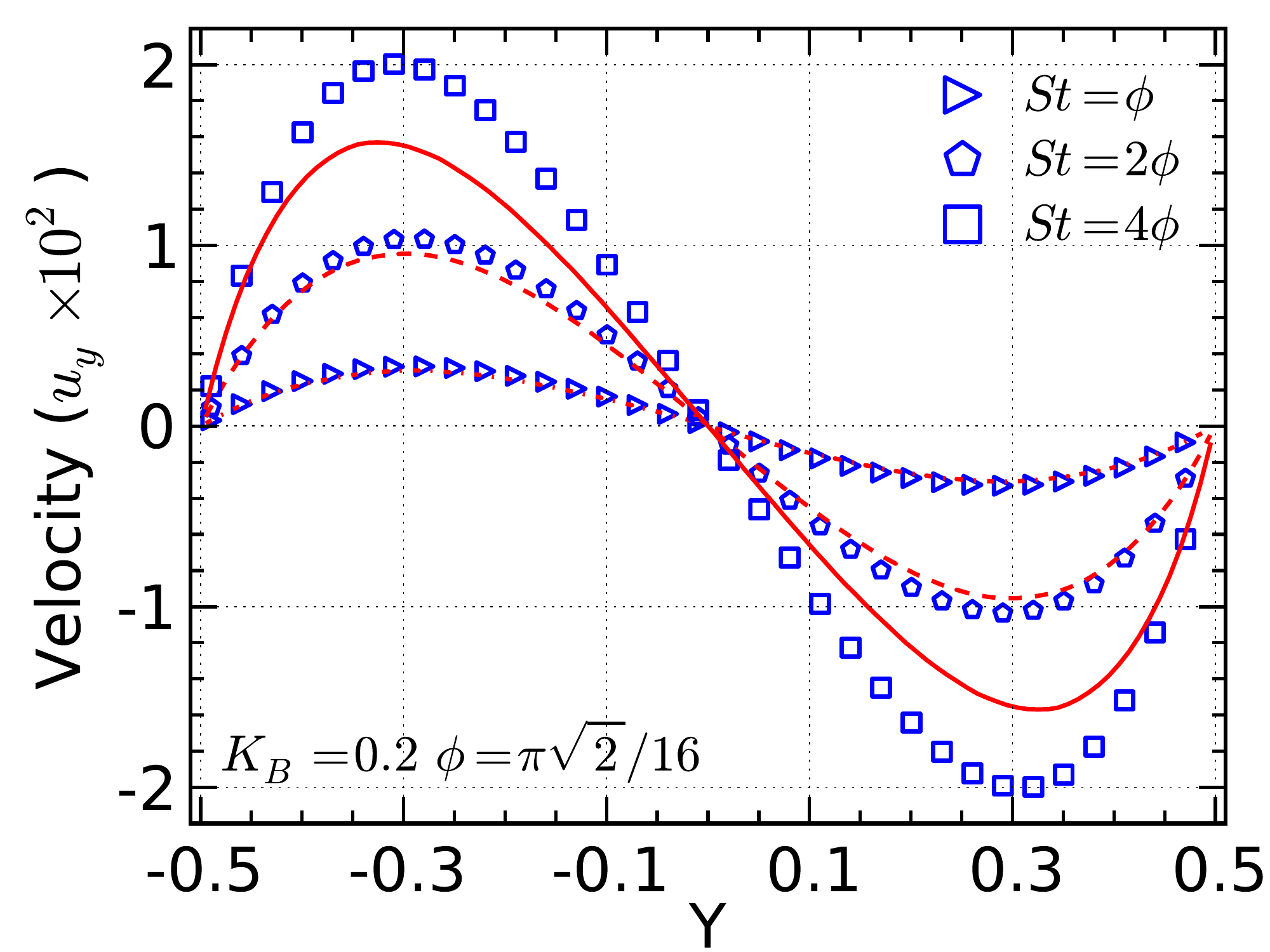}  
\includegraphics[width=0.48 \textwidth,height=0.36 \textwidth]{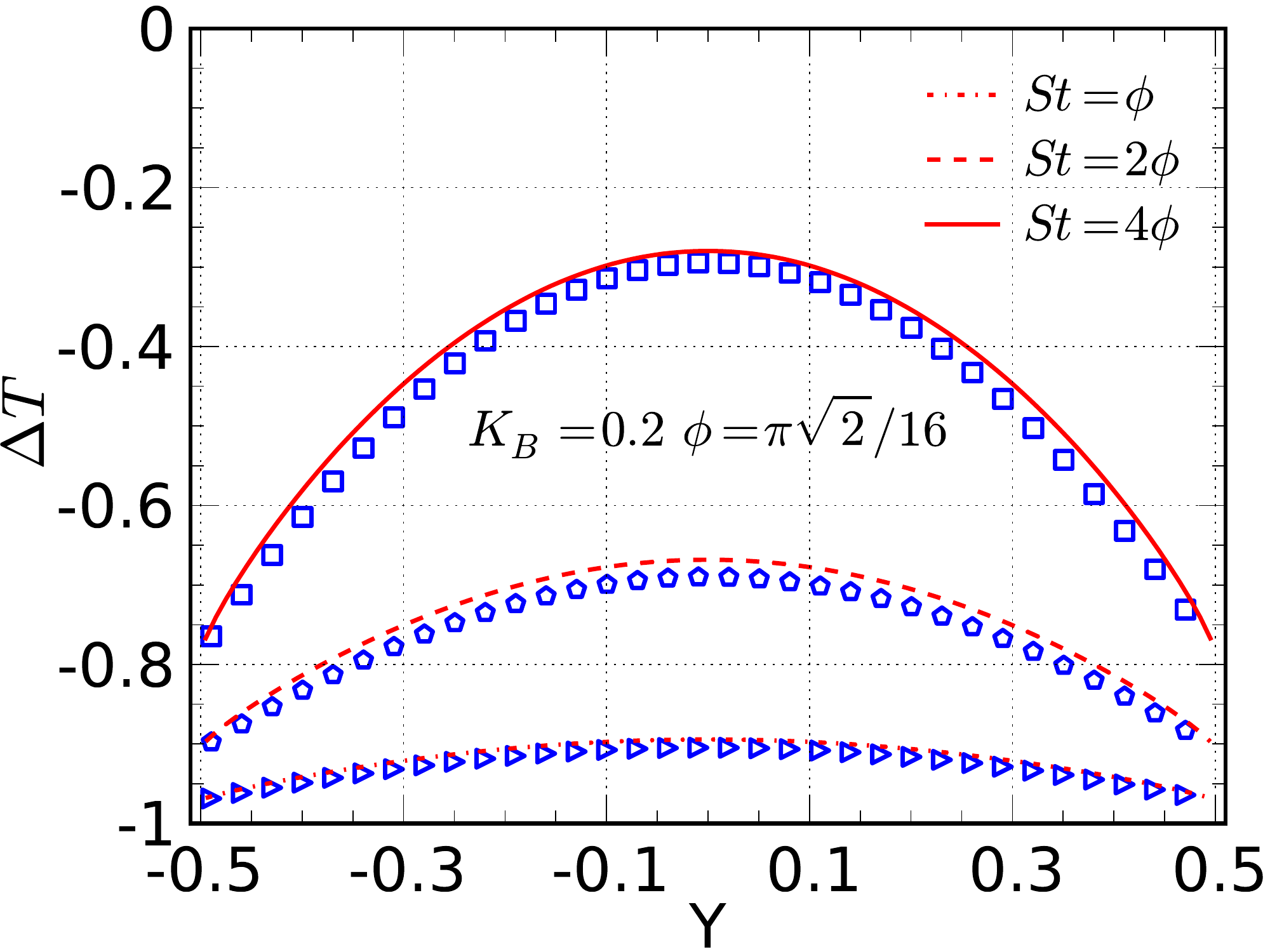}
\caption{The velocity and temperature perturbations for a BGK gas
subject to a sinusoidal heating at $t=3\pi/2$ at different Kn and St. The symbols correspond to the data of the lattice
  model ($Pr=1$) and the lines are the results of the LVDSMC method. \label{ubgkkn02}} 
\end{center}
\end{figure*}

\begin{figure*}
\begin{center}
\includegraphics[width=0.48 \textwidth,height=0.36 \textwidth]{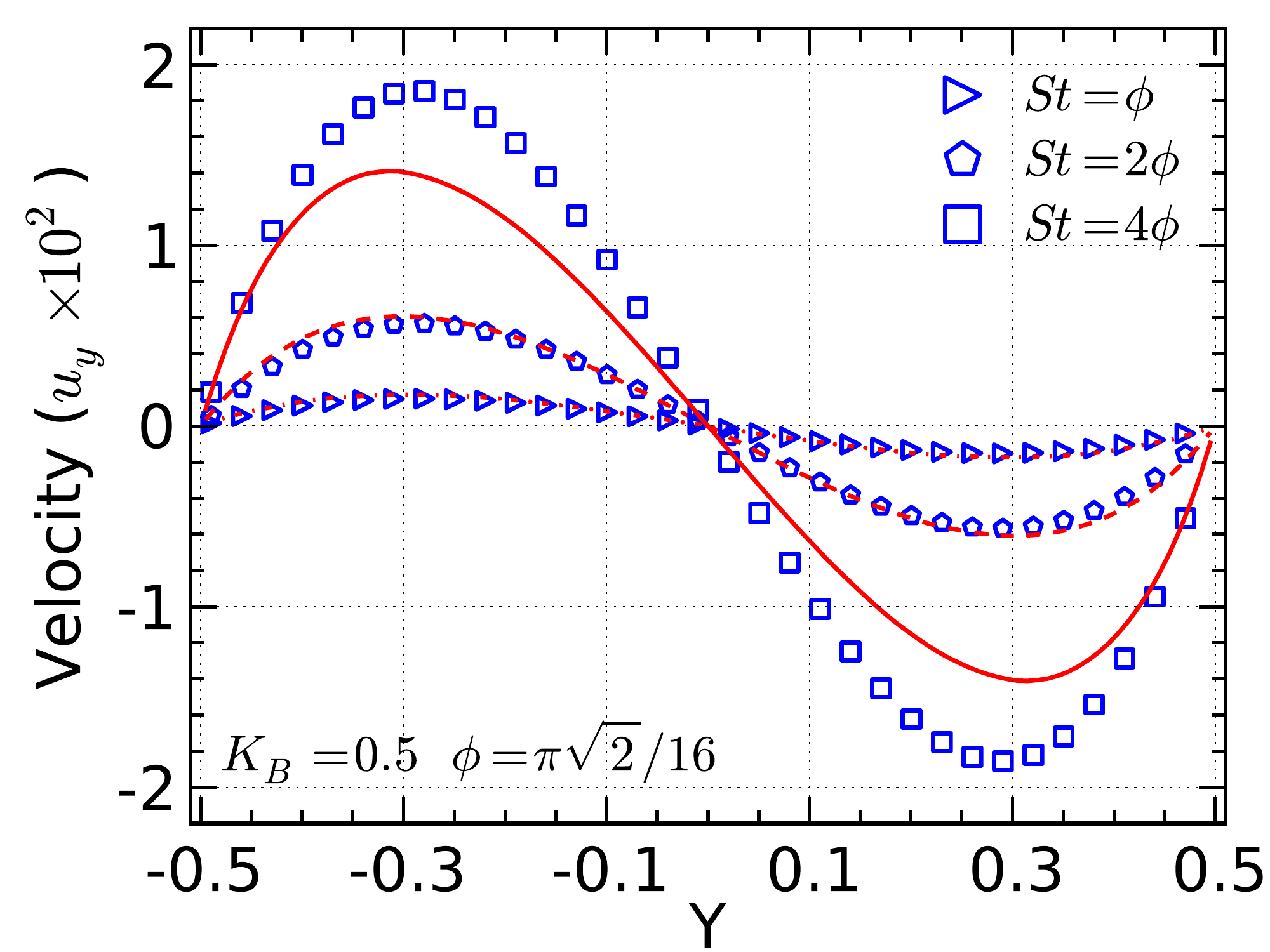}  
\includegraphics[width=0.48 \textwidth,height=0.36 \textwidth]{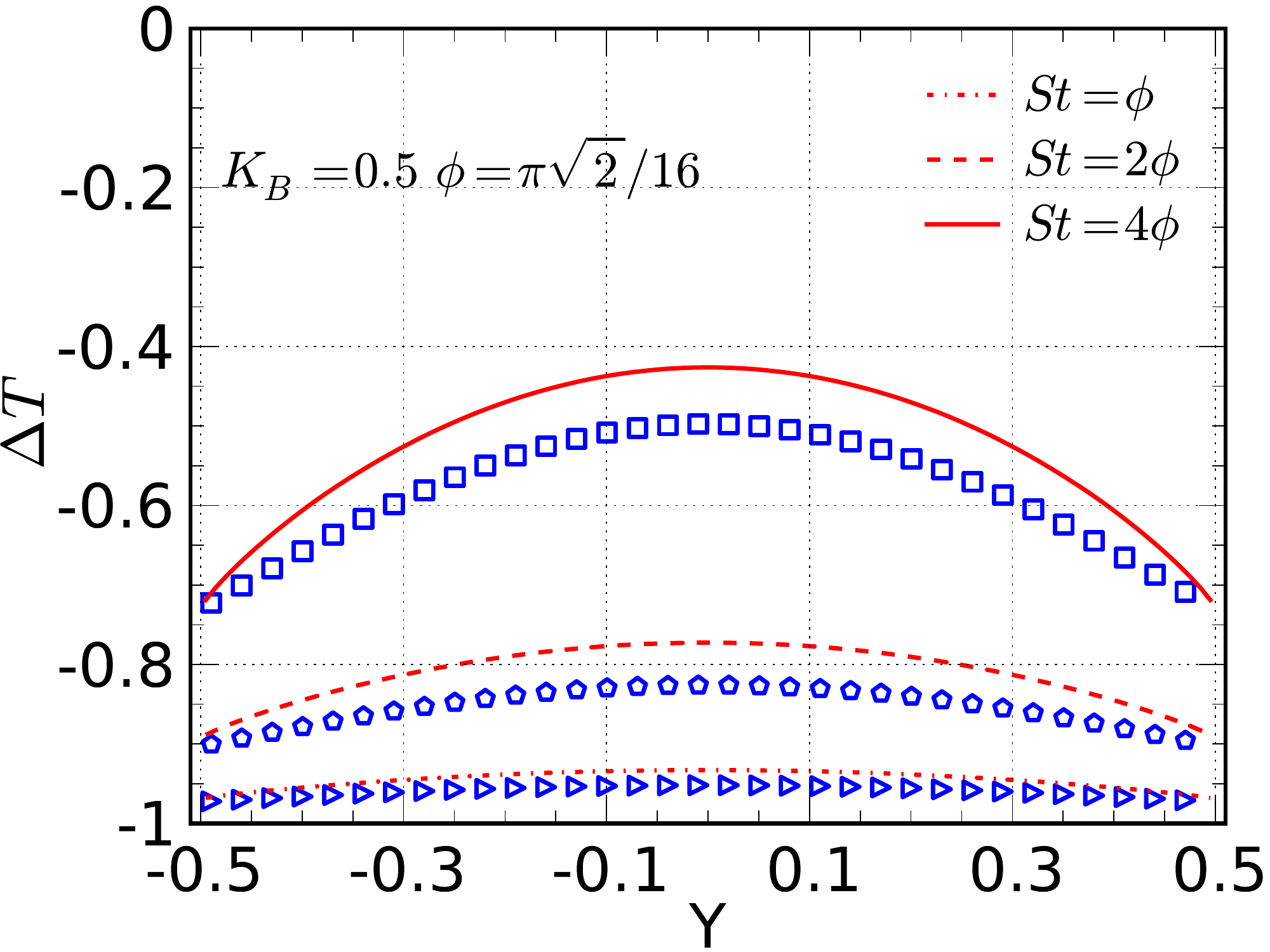}
\caption{The velocity and temperature perturbations for a BGK
  gas subject to a sinusoidal heating at $t=3\pi/2$ at different Kn and St. The symbols correspond to the data of the lattice
  model ($Pr=1$) and the lines are the results of the LVDSMC methods.
\label{ubgkkn05} } 
\end{center}
\end{figure*}

The results for various Strouhal and Knudsen numbers for the hard-sphere and the presented in
Figs.\ref{uhskn005}-\ref{uhskn05}, where both the velocity $u_y$ and
temperature perturbation $\Delta T=T-1$ are normalized by the
amplitude of $F(t)$. Figs.\ref{ubgkkn005}-\ref{ubgkkn05} show results for the BGK model. Symbols denote the lattice model proposed here, whereas
lines denote LVDSMC data.  
At $St=\pi \sqrt{2} /16$, the two sets of results show good agreement, even when $K_E=0.5$ or $K_B=0.5$. With increasing Strouhal number, disprepancies between the two appear and  become
larger. For $St=\pi \sqrt{2}/4$, significant disagreement is observed even for $K_E$ as low as $0.05$. Larger Strouhal numbers lead to stronger rarefaction effects, so this disagreement can be attributed to the moderate discrete velocity set. Overall, with this moderate discrete velocity set, the present model can give reasonable predictions for flows with a  Knudsen number up to $0.5$ and a Strouhal number up to $\pi \sqrt{2}/8$. If highly accurate results are desirable, more discrete velocities are required, leading to higher computational costs.

\subsection{High-Mach number flows}
In this section we present simulation results for high Mach number flows such as Couette flows 
with  $u_w\pm 0.2$ and forced Poiseuille flows with  non-dimensional forcing magnitude $g=0.22$. Although in the 
Couette flows the wall speed is only a factor of 2 larger than the flows examined in section \ref{lowMa}, our focus 
here turns to the resulting temperature field that, despite the small temperature differences involved, reveals useful information.

\begin{figure*}
\begin{center}
\includegraphics[width=0.48 \textwidth,height=0.36 \textwidth]{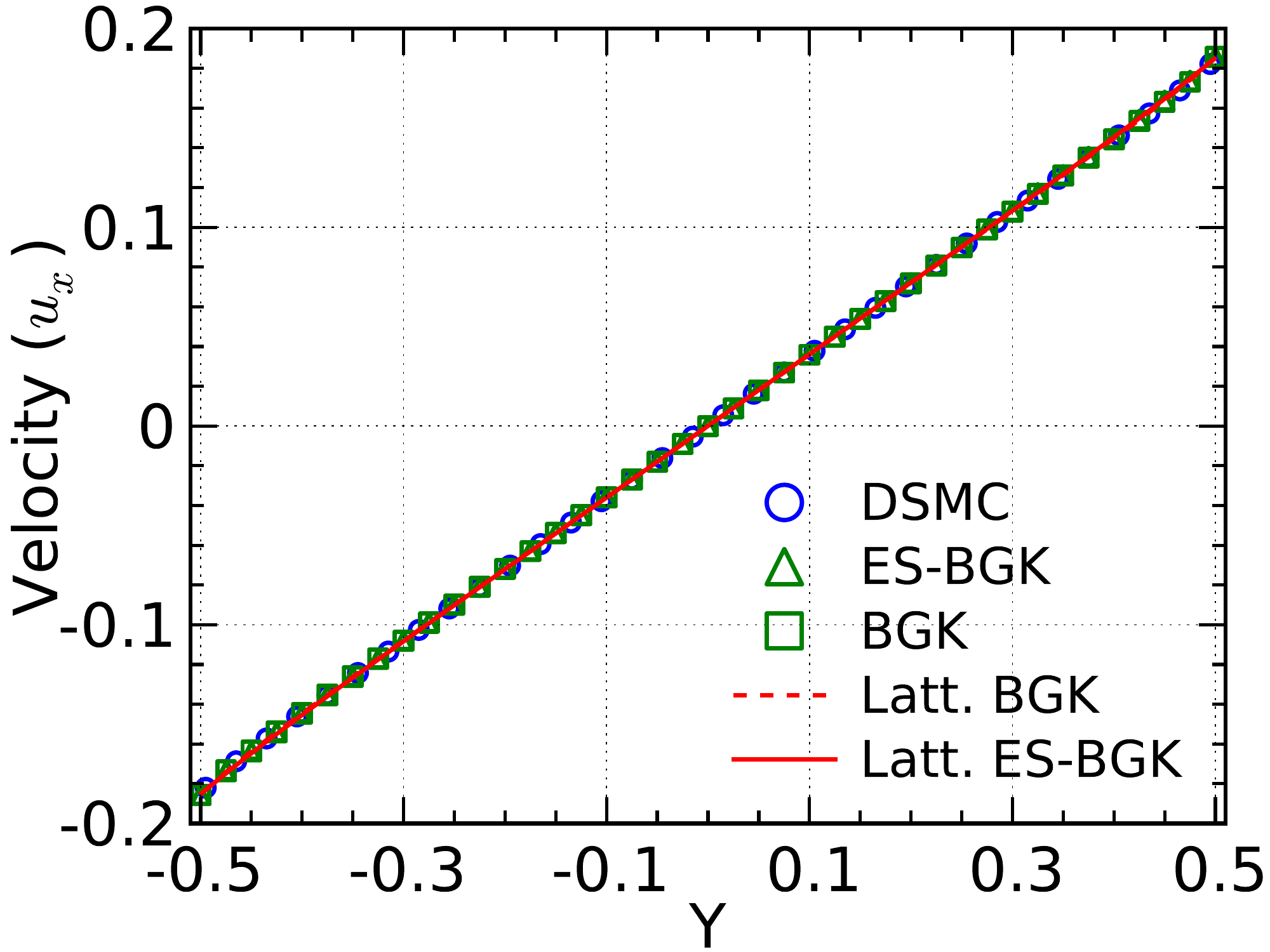}  
\includegraphics[width=0.48 \textwidth,height=0.36 \textwidth]{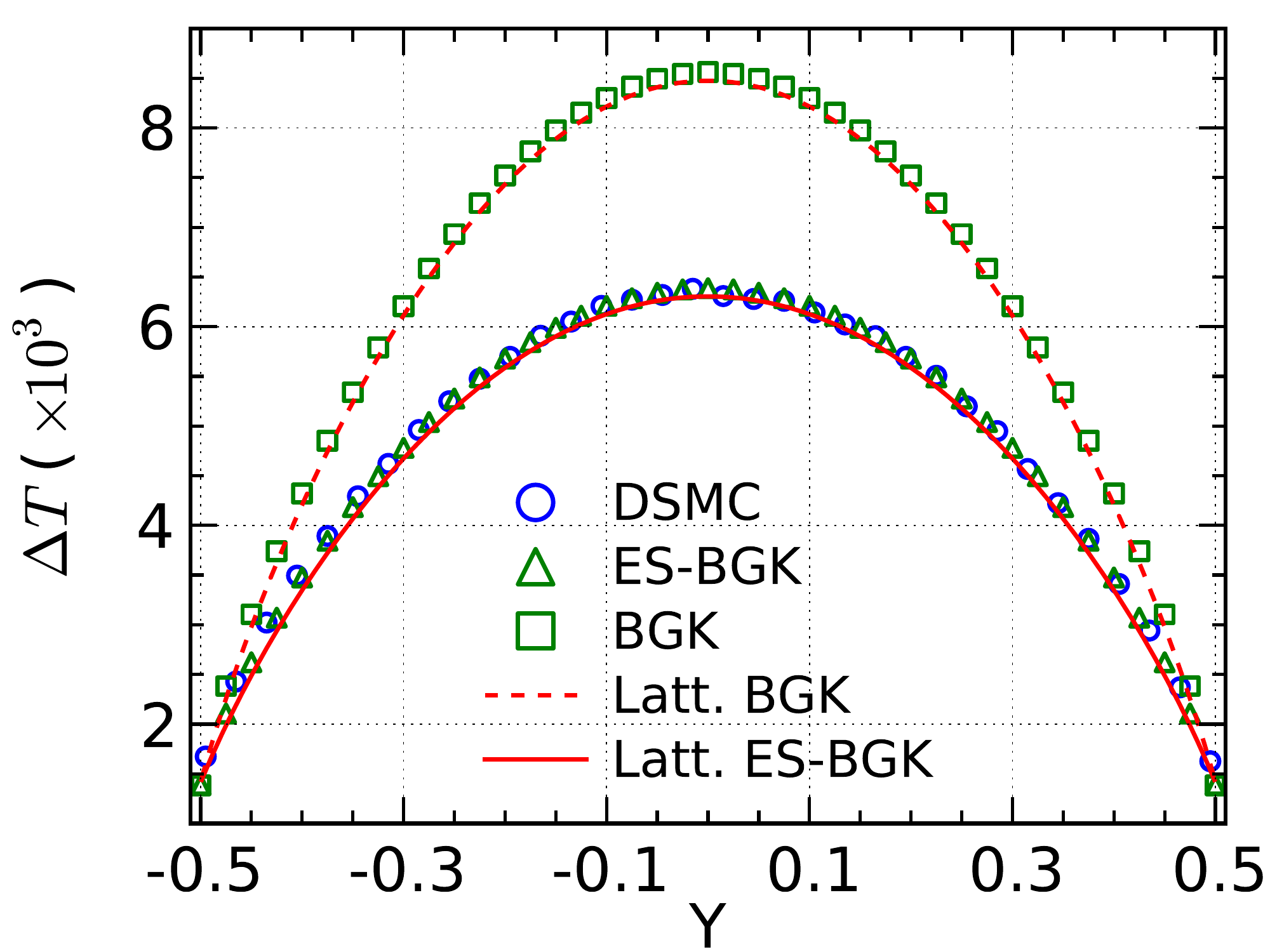}
\caption{The velocity and temperature profiles for the Couette flow at
  $K_D=0.05$, where $\Delta T=T-T_w$. The wall velocities are $u_w=\pm 0.2$. \label{prandtl}}
\end{center}
\end{figure*}

Fig.~\ref{prandtl} shows a comparison between our lattice model, DSMC and numerical 
solution of the ES-BGK equation for a Couette flow at $K_D=0.05$. The comparison reveals 
that at these small $K_D$, the lattice model can capture both kinetic (e.g. slip/temperature jump) and non-equilibrium effects 
quite accurately. For comparison, the results of the BGK equation 
and its lattice version  ($b=0$) are also presented. These two
models predict a temperature maximum that is $30\%$ higher than
the hard-sphere result.

\begin{figure*}
\begin{center}
\includegraphics[width=0.48 \textwidth,height=0.36 \textwidth]{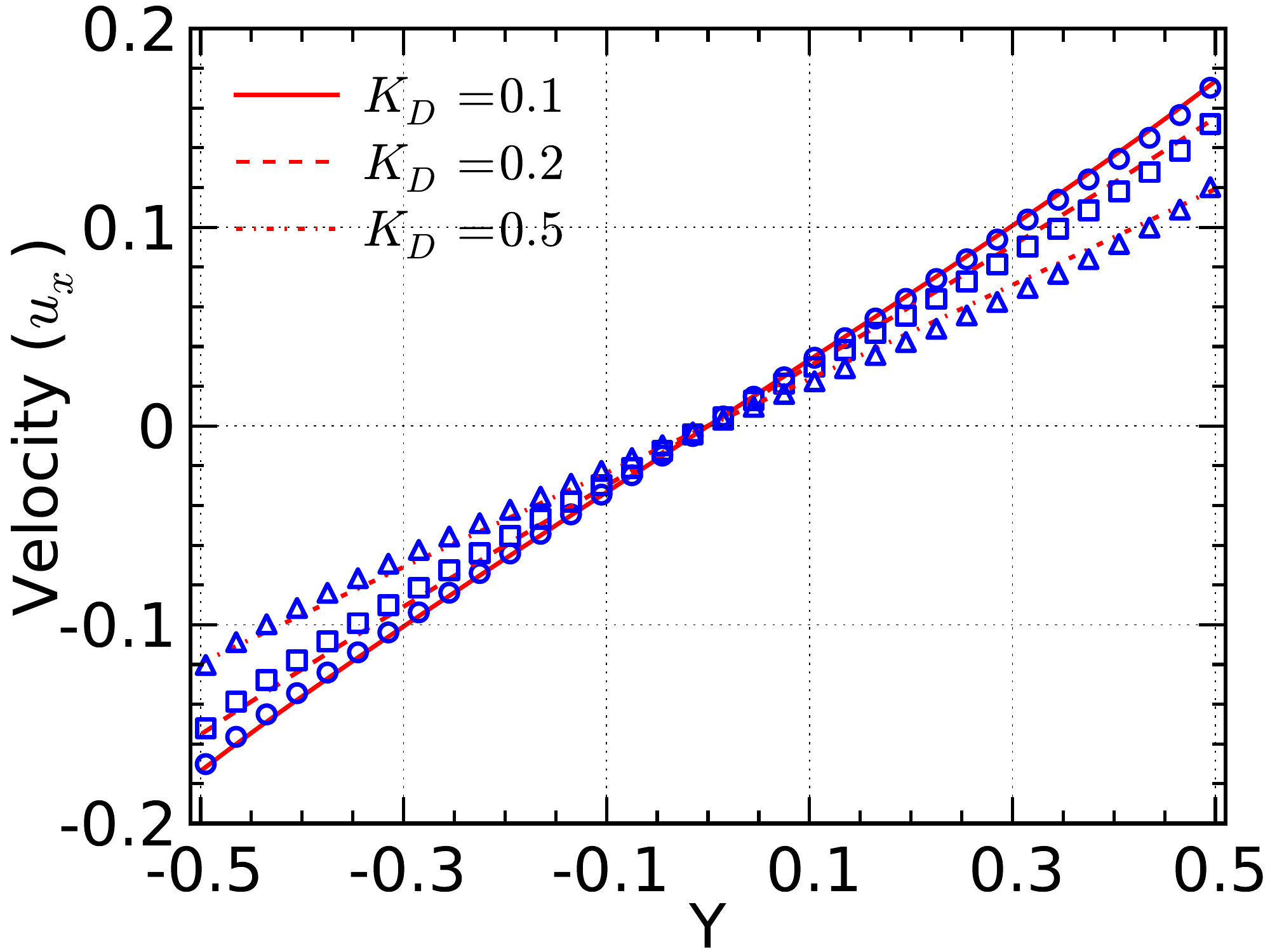}  
\includegraphics[width=0.48 \textwidth,height=0.36
\textwidth]{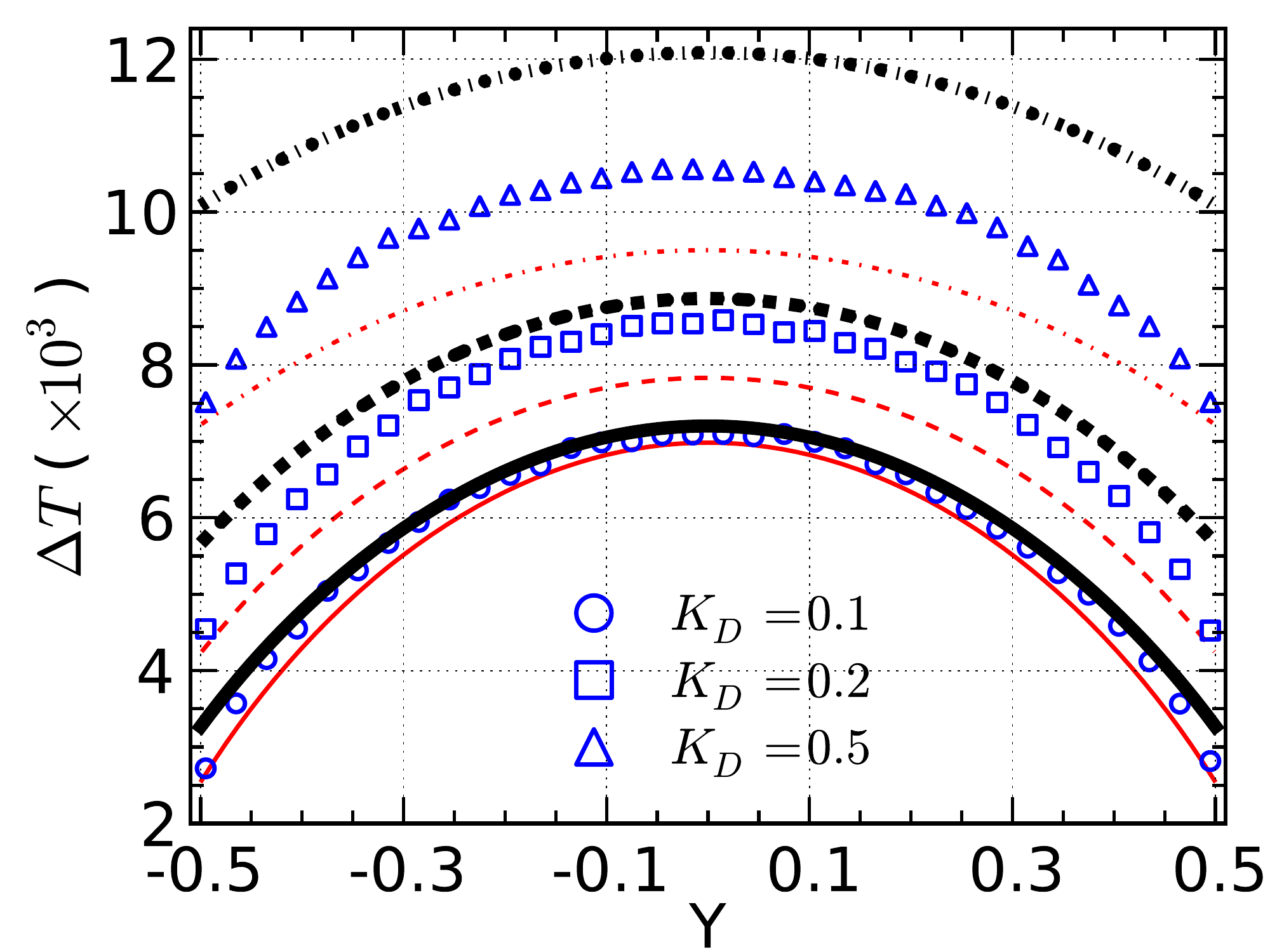}
\caption{The velocity and temperature profiles ($\Delta T=T-T_w$)  for Couette
flow of a hard-sphere gas at $K_D=0.1$, $0.2$ and $0.5$. The symbols denote the DSMC data
 and the red lines represent the lattice ES-BGK
  results. The wall velocities are $u_w=\pm 0.2$. For
  further comparison, the temperature profiles predicted by
  the R13 model \citep{taheri:017102} are also presented
  with the black lines where the line styles same to those of
  the lattice
  ES-BGK results are used to distinguish the Knudsen
  numbers.    \label{couetteut}} 
\end{center}
\end{figure*}

\begin{figure*}
\begin{center}
\includegraphics[width=0.48 \textwidth,height=0.36 \textwidth]{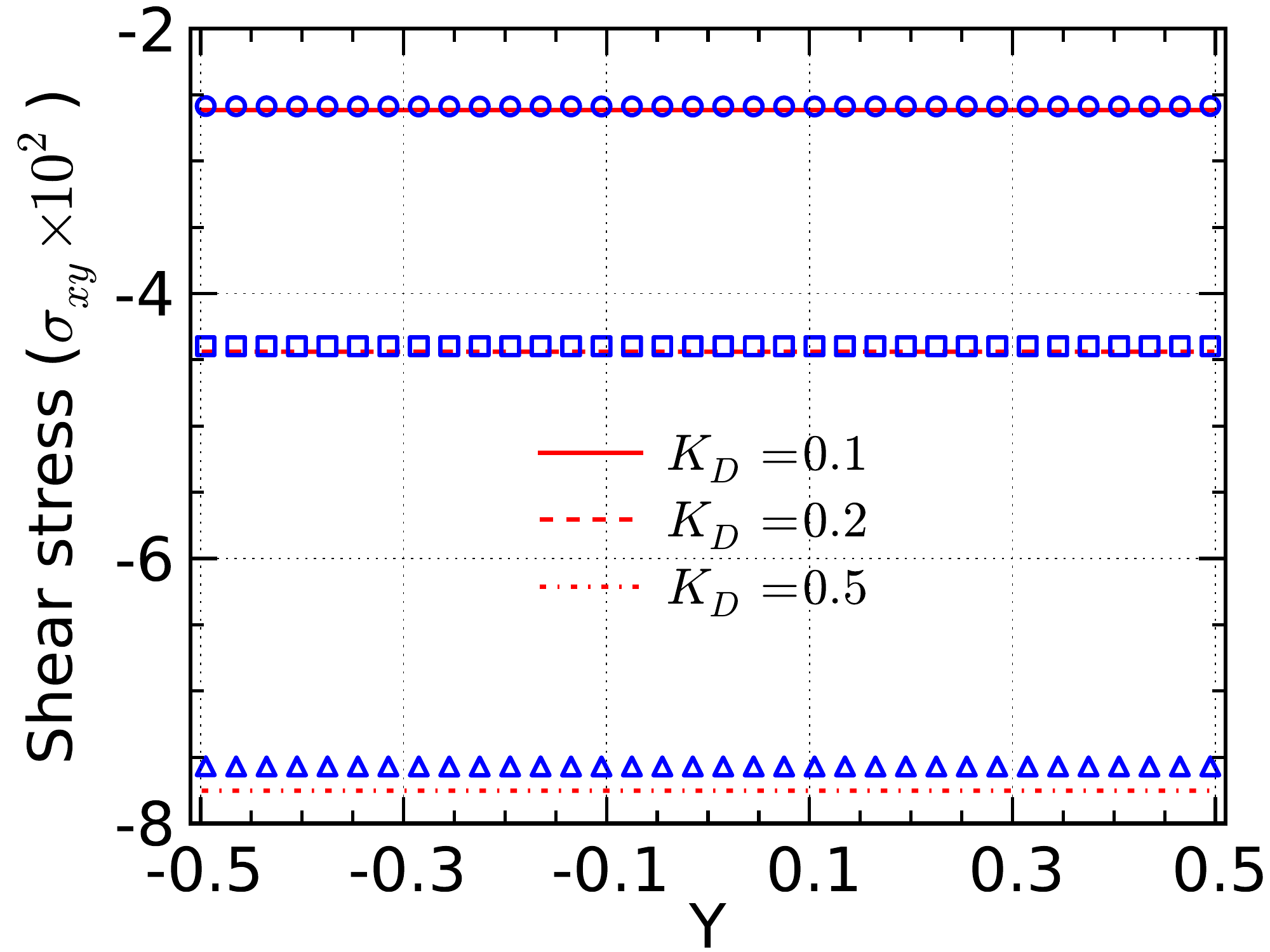}
\includegraphics[width=0.48 \textwidth,height=0.36 \textwidth]{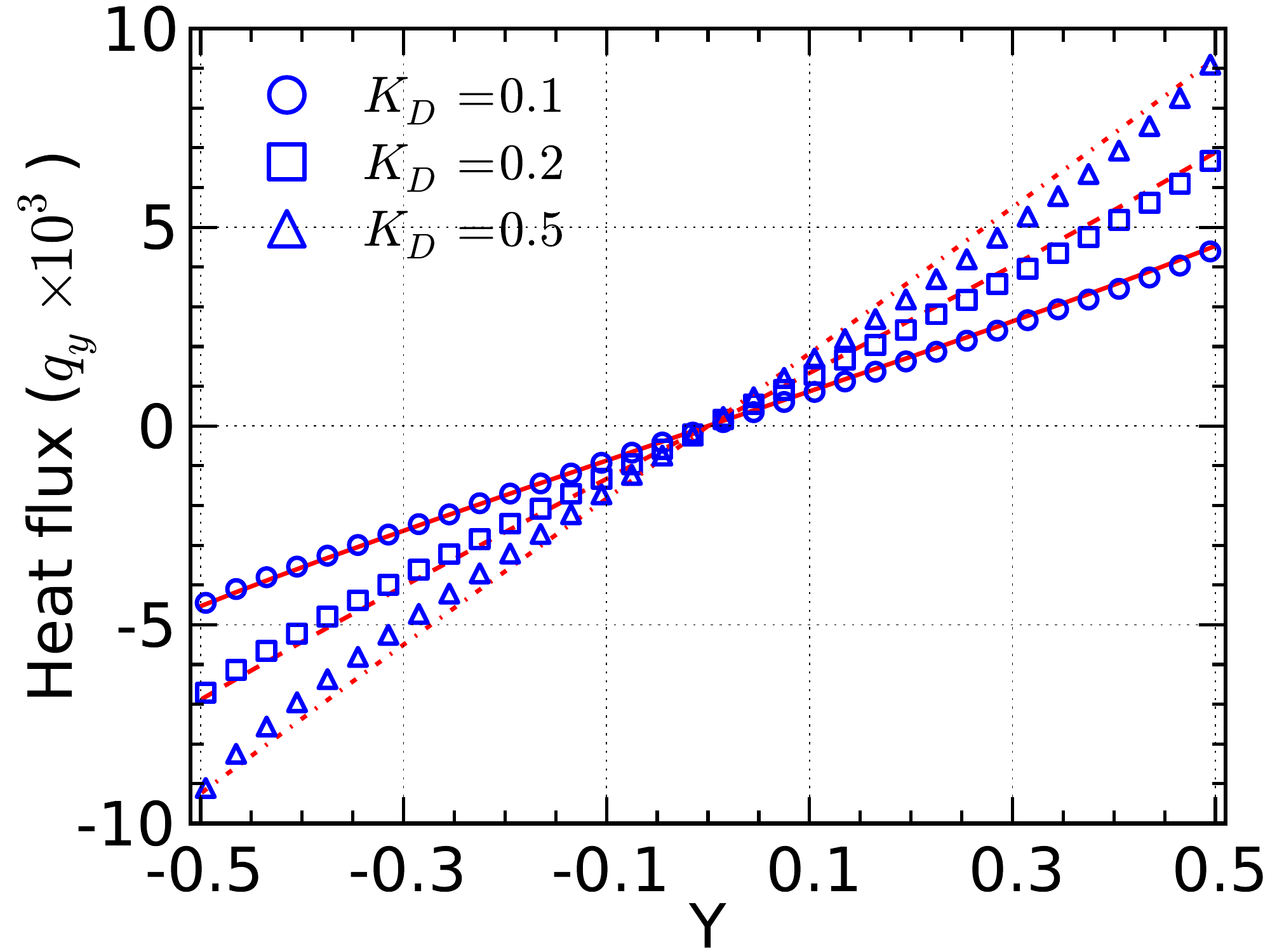}
\caption{The shear stress ($\sigma_{xy}$) and heat flux ($q_y$)
  profiles for Couette flow of
  a hard-sphere gas at $K_D=0.1$, $0.2$ and $0.5$. The symbols denote DSMC data
  and the lines represent the lattice ES-BGK results. The wall velocities are $u_w=\pm 0.2$. \label{couettesh}} 
\end{center}
\end{figure*}

\begin{figure*}
\begin{center}
\includegraphics[width=0.48 \textwidth,height=0.36 \textwidth]{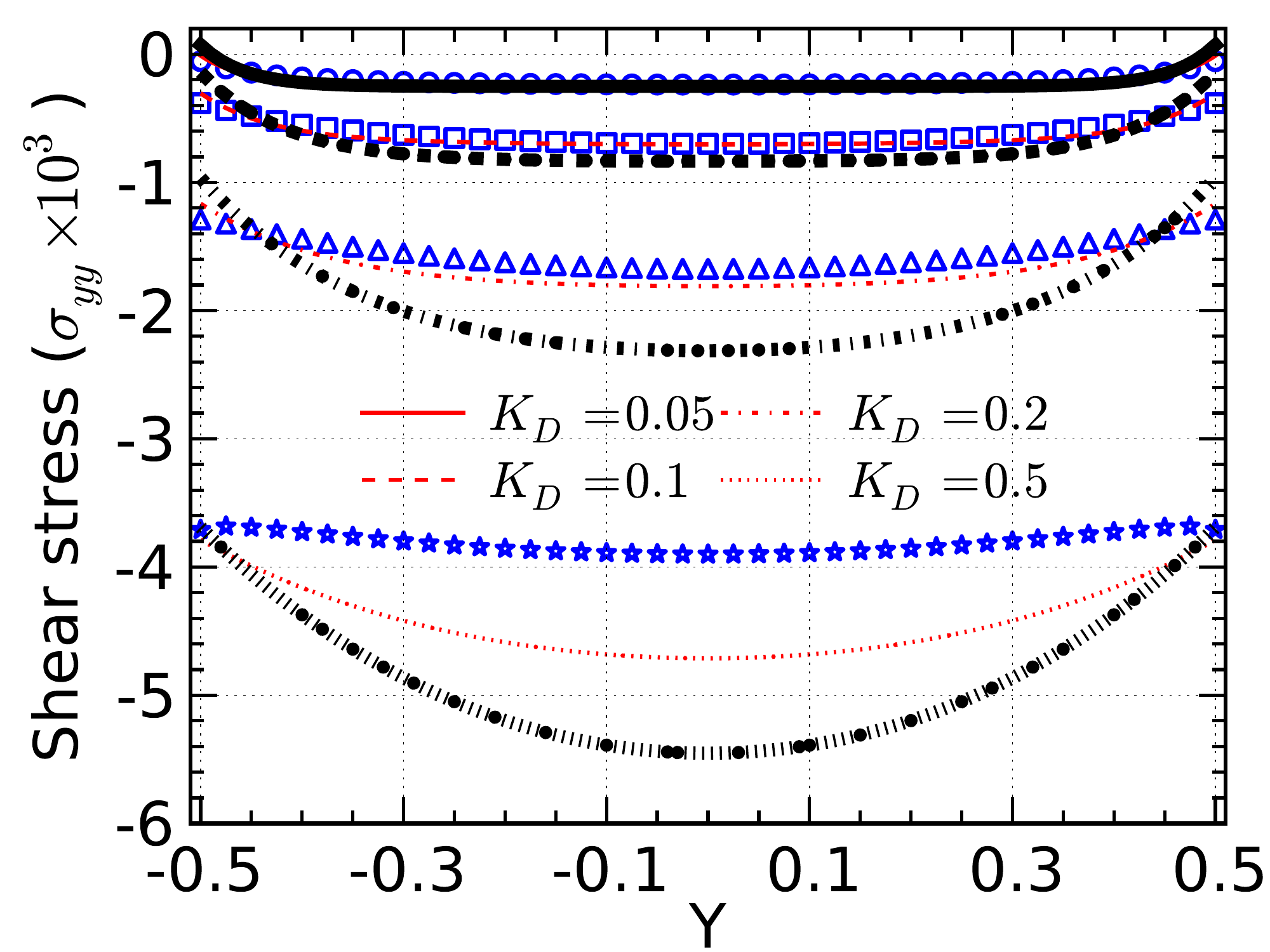}
\includegraphics[width=0.48 \textwidth,height=0.36 \textwidth]{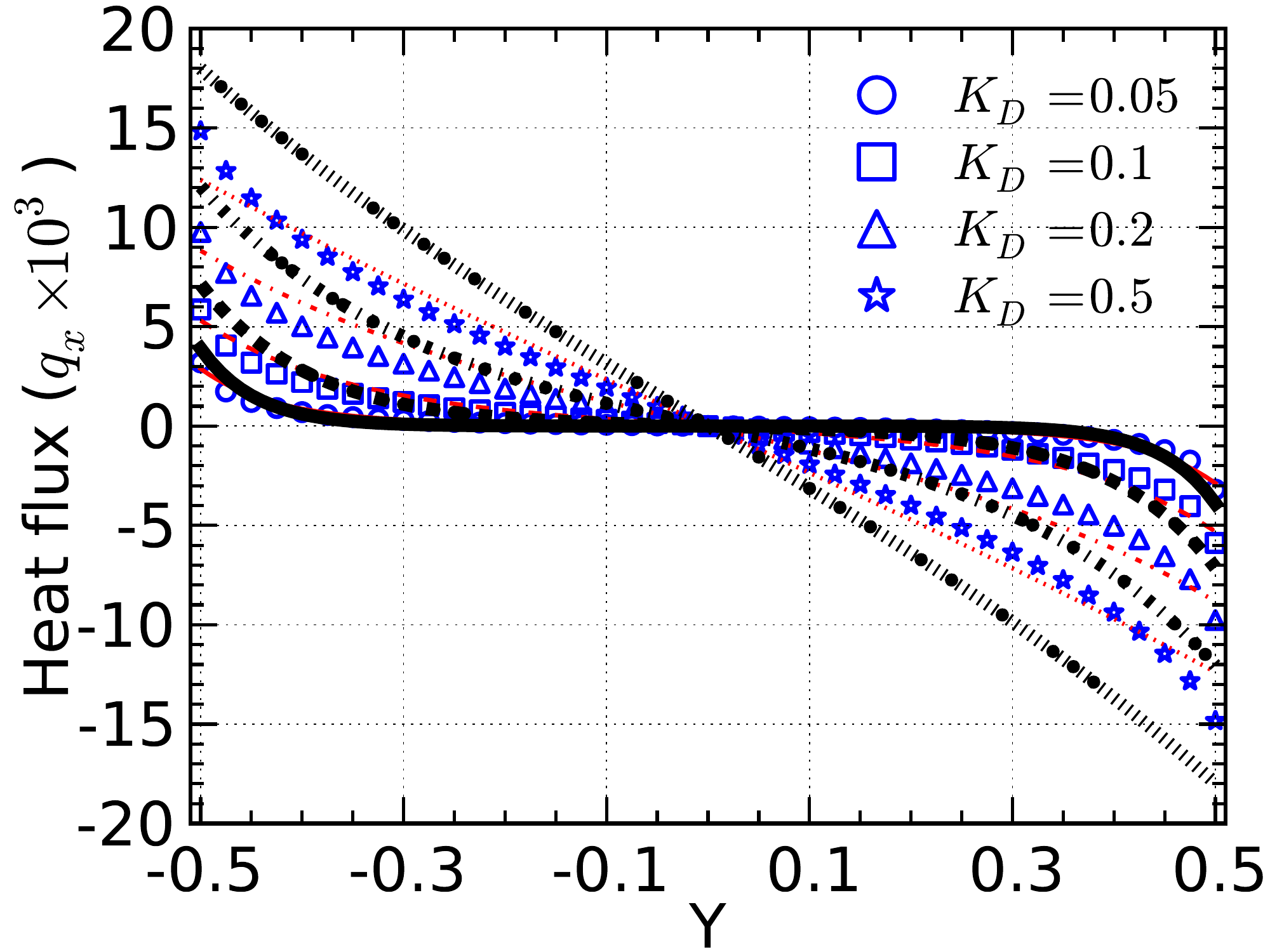}
\caption{The shear stress ($\sigma_{yy}$) and heat flux ($q_x$) profiles for Couette flow of
  a hard-sphere gas at $K_D=0.1$, $0.2$ and $0.5$. The
  symbols denote direct solution of the ES-BGK equation, the
  red lines represent the lattice ES-BGK results and the
  black lines are those of the R13 model \citep{taheri:017102}. The Knudsen numbers for the
 R13 model are denoted using the same line styles as the lattice ES-BGK results. The wall velocities are $u_w=\pm 0.2$.\label{couetteqxsyy}} 
\end{center}
\end{figure*}
\begin{figure*}
\begin{center}
\includegraphics[width=0.48 \textwidth,height=0.36 \textwidth]{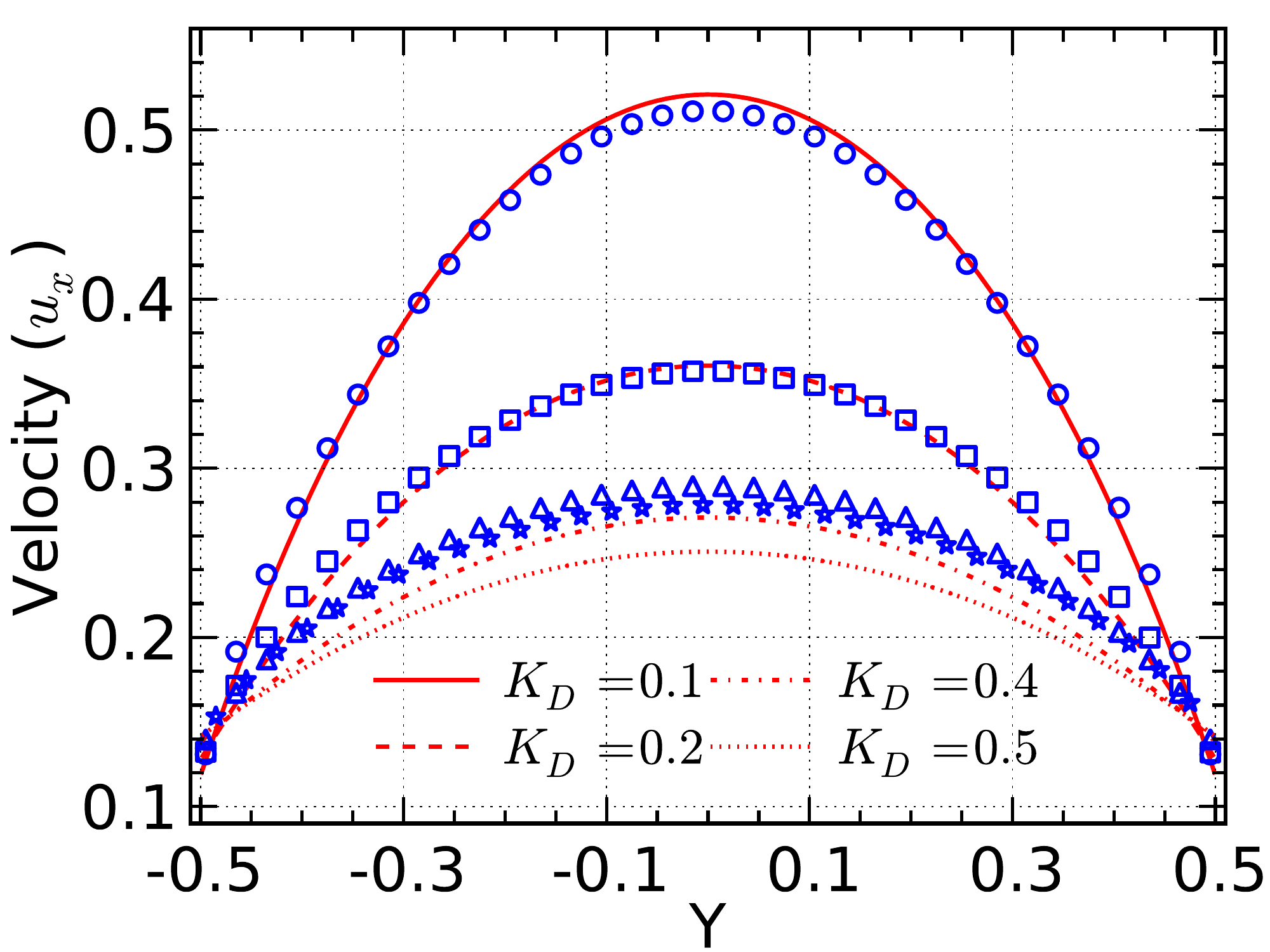}  
\includegraphics[width=0.48 \textwidth,height=0.36 \textwidth]{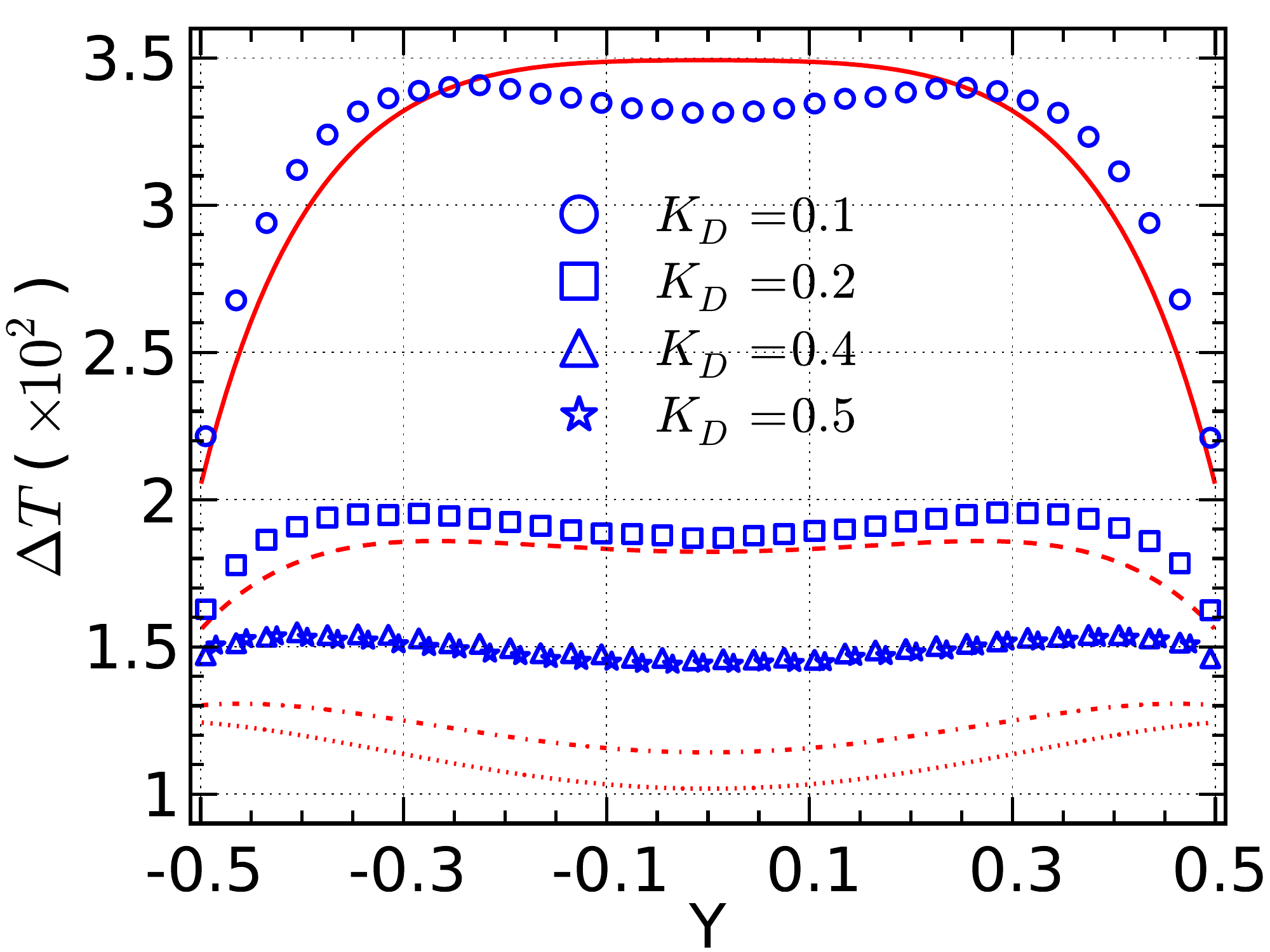}
\caption{The velocity and temperature profiles ($\Delta T=T-T_w$) for the force-driven
Poiseuille flow of a hard-sphere gas at $K_D=0.1$, $0.2$, $0.4$ and $0.5$. The symbols denote DSMC data
  and the lines represent the lattice ES-BGK results. 
\label{posiut} }
\end{center}
\end{figure*}

\begin{figure*}
\begin{center}
\includegraphics[width=0.48 \textwidth,height=0.36 \textwidth]{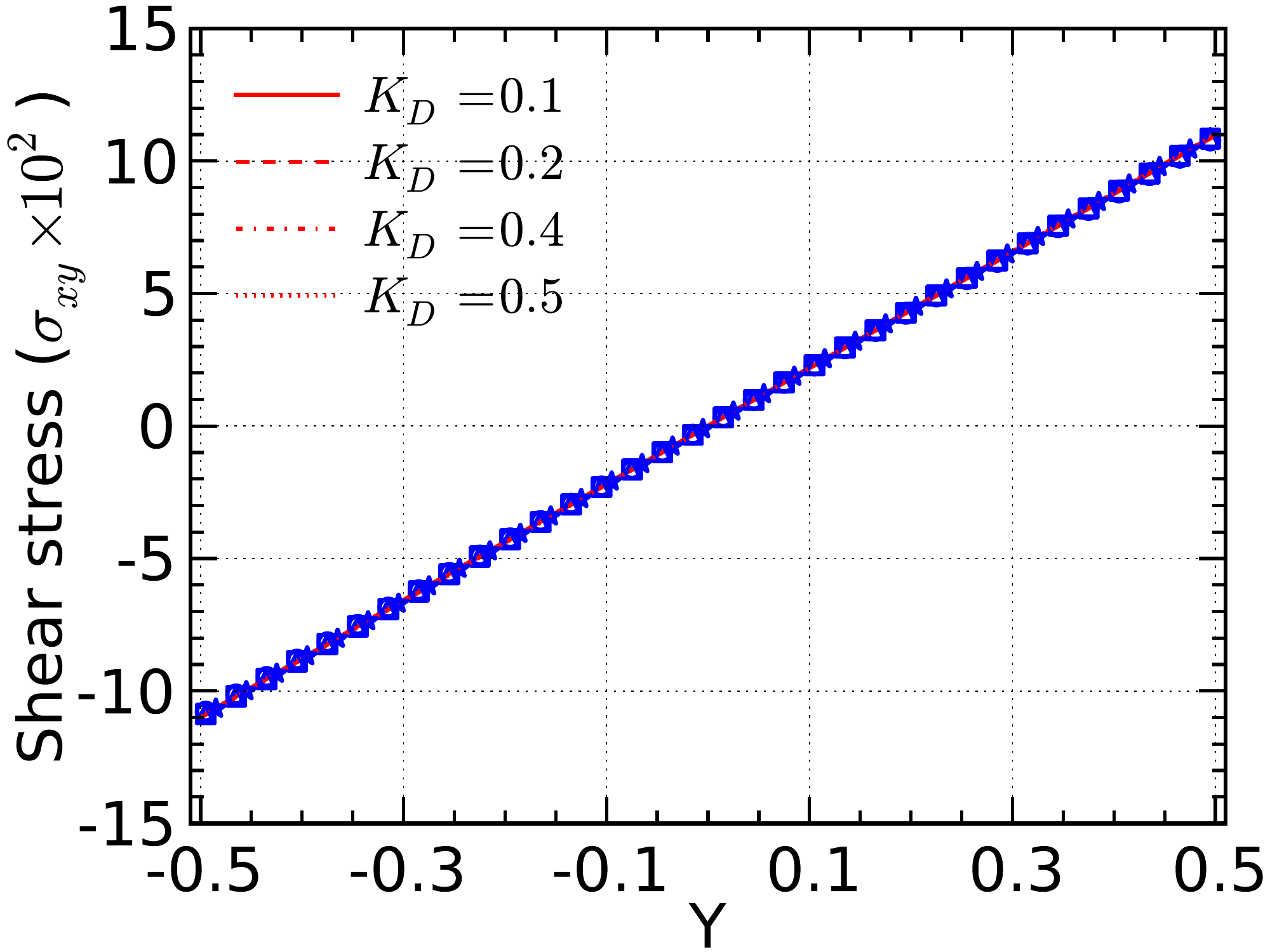}
\includegraphics[width=0.48 \textwidth,height=0.36 \textwidth]{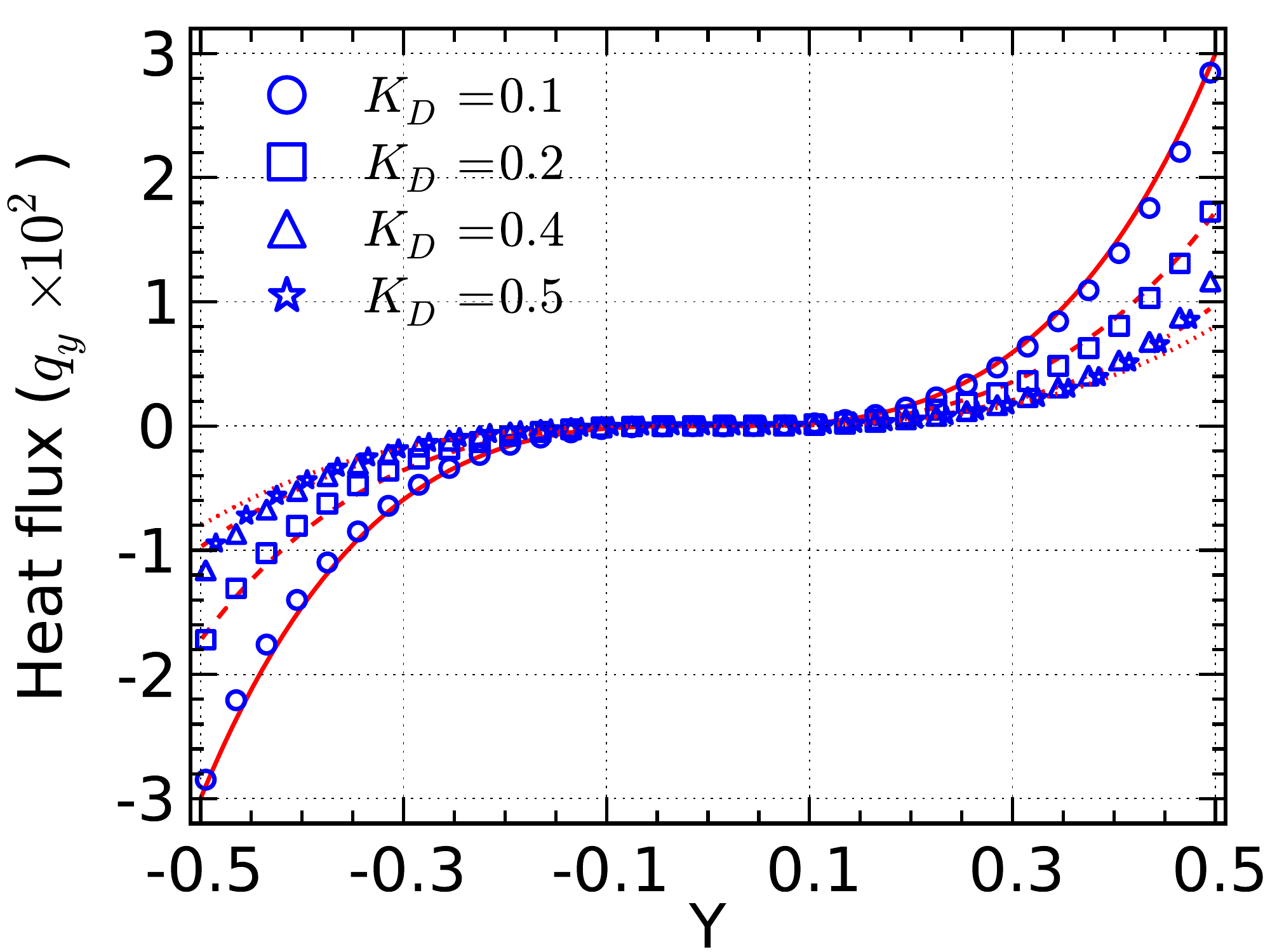}
\caption{The shear stress and heat flux profiles for the force-driven Poiseuille flow of
  a hard-sphere gas at $K_D=0.1$, $0.2$, $0.4$ and $0.5$. The symbols denote DSMC data
  and the lines represent the lattice ES-BGK results. \label{posish}}
\end{center}
\end{figure*}

\begin{figure*}
\begin{center}
\includegraphics[width=0.48 \textwidth,height=0.36 \textwidth]{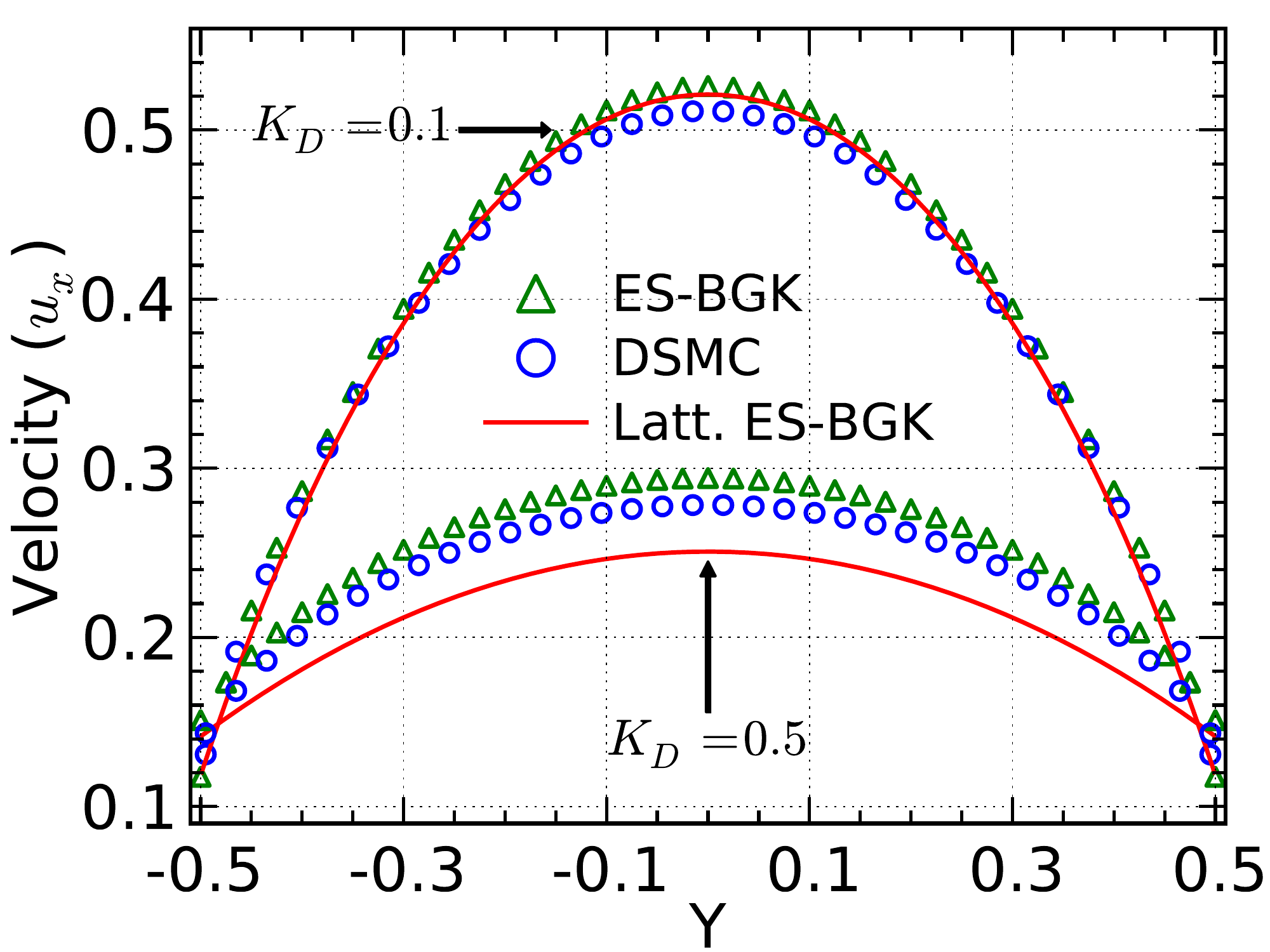}
\includegraphics[width=0.48 \textwidth,height=0.36 \textwidth]{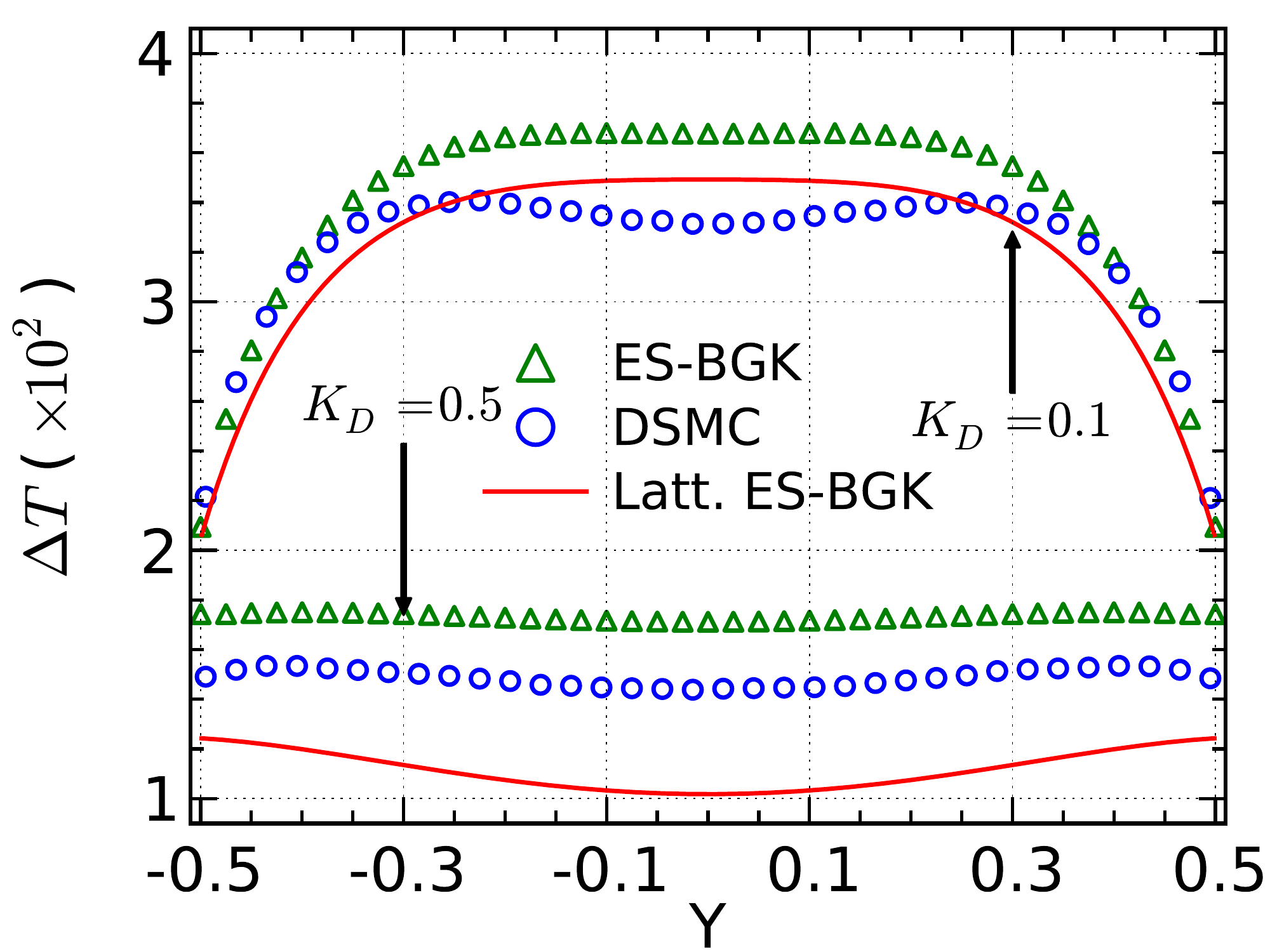}
\caption{Comparisons of the velocity and temperature profiles ($\Delta T=T-T_w$)
for the force-driven flow  among the ES-BGK equation, lattice ES-BGK model and DSMC method at $K_D=0.1$ and
$0.5$. \label{posiesut}}
\end{center}
\end{figure*}

\begin{figure*}
\begin{center}
\includegraphics[width=0.48 \textwidth,height=0.36 \textwidth]{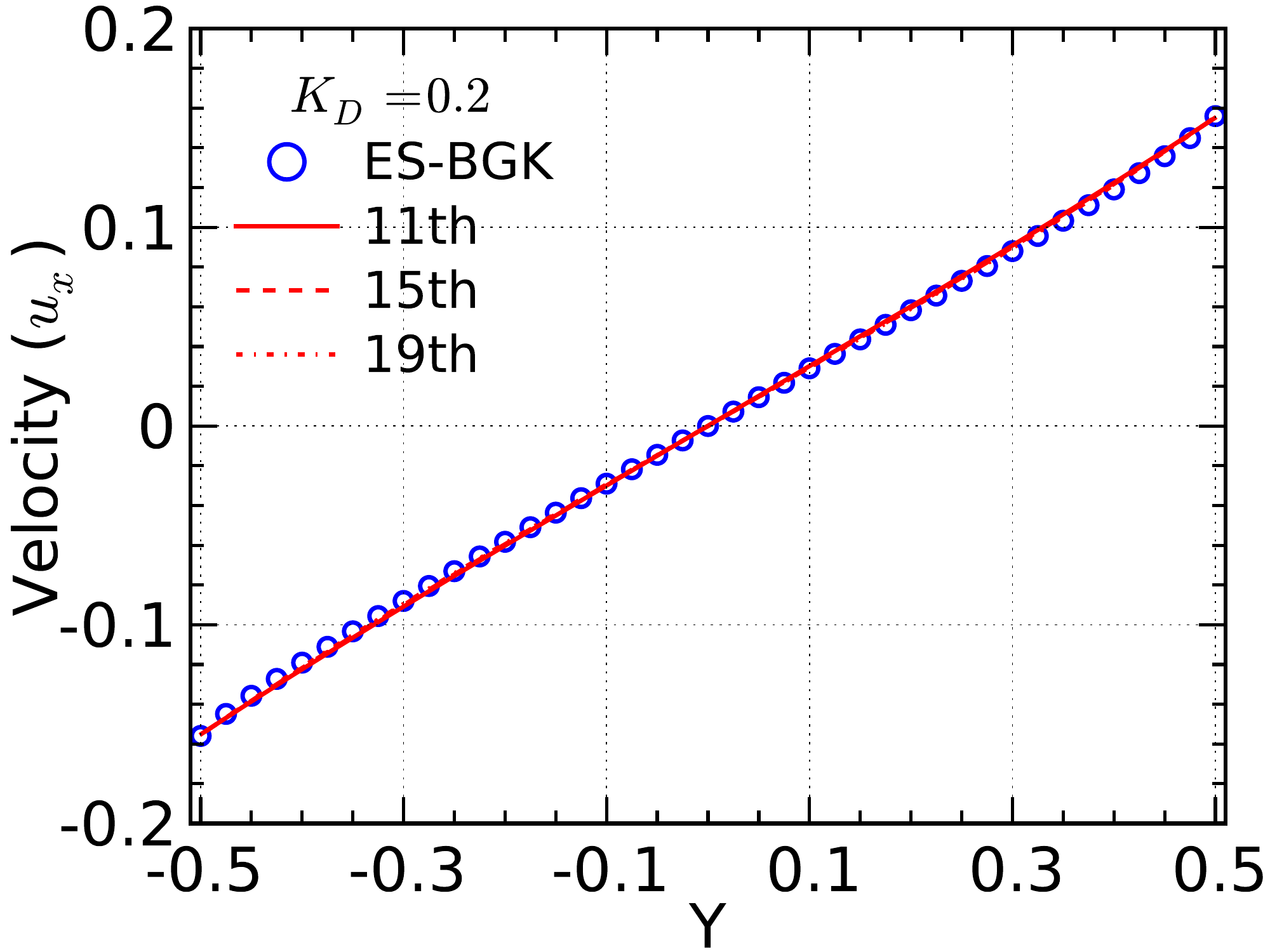}  
\includegraphics[width=0.48 \textwidth,height=0.36
\textwidth]{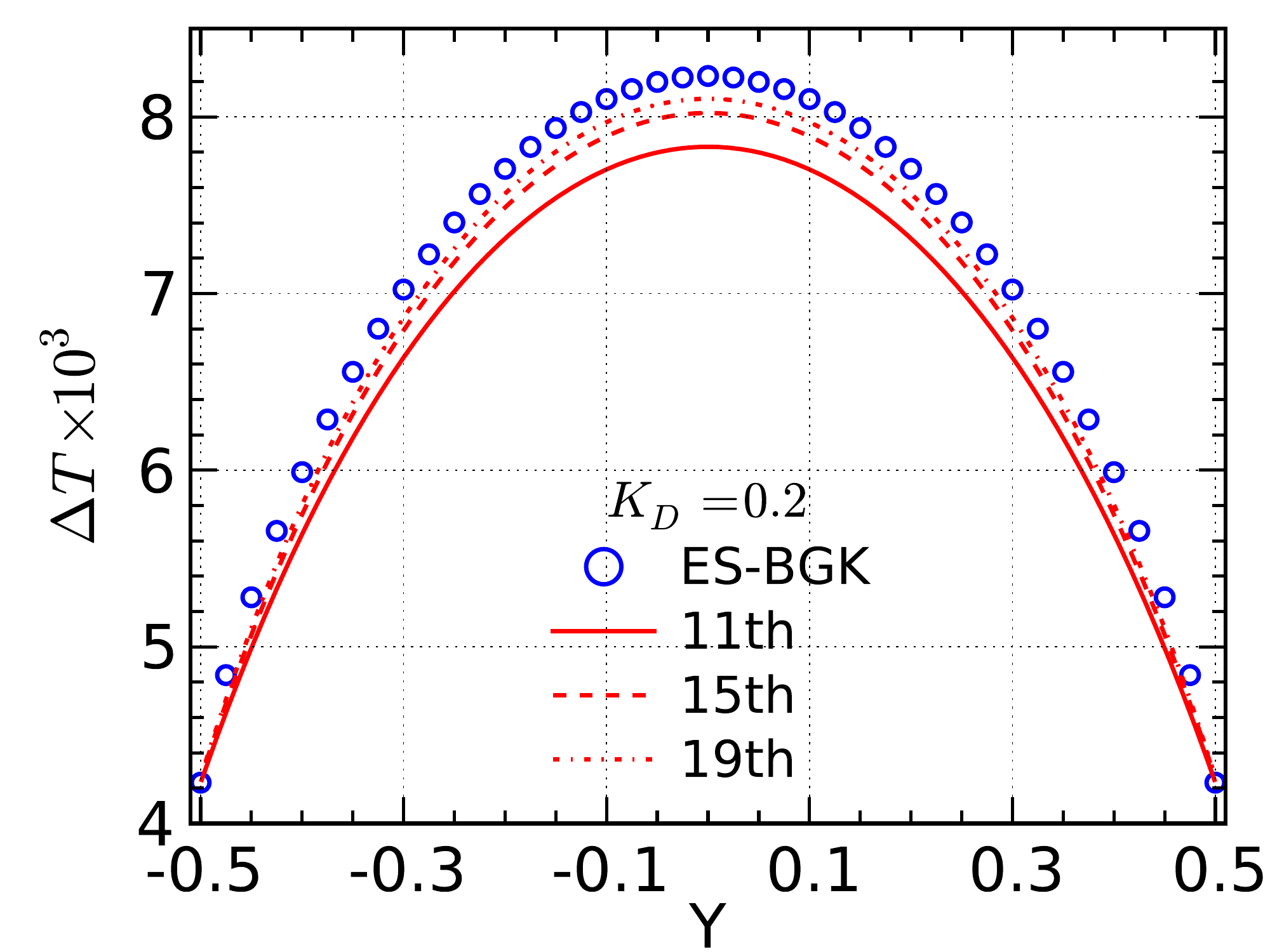}
\caption{Convergence comparison for a Couette flow at
  $K_D=0.2$ with wall velocities $u_w=\pm 0.2$. Lines denote the lattice model data, 
and the adopted discrete velocity order as denoted in the legend. Improved accuracy is observed as the number of discrete velocities is increased.
 \label{cvg}} 
\end{center}
\end{figure*}


Results for Couette flows in the transition regime are presented in
Figs.~\ref{couetteut} and~\ref{couettesh}. Overall, our model predictions 
for the velocity, shear stress and heat flux are close to those of DSMC 
even for $K_D=0.5$. However, our predictions of the viscous-heating-induced 
temperature  field start to deviate from the DSMC results at $K_D=0.2$. 
Remarkably, the heat flux predictions of the lattice model are still 
in excellent agreement with the DSMC data for $K_D$ as large
as 0.5.

To further validate the model accuracy beyond the
  NS level, the profiles of shear stress
  $\sigma_{yy}$ and stream-wise heat flux $q_x$ are
  presented in Fig.(\ref{couetteqxsyy}), which are both zero
  at the NS level of approximation. However, our lattice
  model can give reasonable predictions for $\sigma_{yy}$ up
  to $K_D=0.2$ which is consistent with the temperature
  prediction. For the heat flux $q_x$, we have similar
  observation, which further confirms that our model
  can describe kinetic effects beyond the NS level
  with the chosen moderate 11-th order quadrature, which is
  consistent with the discussion in
  Sec.\ref{remarkaccuracy}. It is also interesting to compare
  with the predictions of the R13 model, which is supposed to give a stable set of transport equations of the super-Burnett order \citep{taheri:017102}. Fig.(\ref{couetteqxsyy}) shows that the R13 model and our lattice model give similar results.
 
The simulation results for Poiseuille flows at four Knudsen numbers 
are depicted in Figs.~\ref{posiut} and~\ref{posish}. 
This flow is more numerically challenging compared to the
Couette flow.  Even for $K_D$ as small as 0.1, the temperature profiles show
significant difference between the lattice model and DSMC results, 
even though the bimodal temperature distribution
~\citep{Mansour1997} can be qualitatively captured by the lattice model 
at  $K_D=0.2$.
As the Knudsen number increases further, the difference becomes
even larger, though the temperature minimum at the center of the
computational domain can be predicted qualitatively. 
Remarkably, but also consistently with the Couette flow case, 
the heat flux profiles show good agreement with the DSMC data despite 
the more significant discrepancies in the temperature profiles.

The inability of our model to predict accurate viscous heating may be caused by two factors. First, as shown in the derivation process,
the lattice model approaches the corresponding kinetic model via the moment
representation with increasing order of Gauss-Hermite quadrature. 
However, as discussed in Sec.\ref{remarkaccuracy}, the
chosen moderate discrete velocity set is only expected to
approach the Burnett level, which may not be sufficient to capture
all kinetic effects in the presence of a non-negligible Mach
number. This is verified in Fig.~\ref{cvg} where its is
shown that enriching the discrete velocity set can improve the
model accuracy.

In addition to the error due to a finite number of discrete velocities,  
the ES-BGK equation itself is a model which may be failing in these flows. 
In fact, as shown by some studies \citep[e.g.,][]{Gallis2011}, the ES-BGK equation tends to produce
inaccurate velocity distribution functions.  
The possibility that the ES-BGK model is itself contributing to the error can be confirmed by the results presented in 
Fig.~\ref{posiesut}. At $K_D=0.5$ the lattice model result shows a significant deviation from the numerical solution of the ES-BGK model. 
Meanwhile, the temperature profile from numerical solution of the ES-BGK equation itself shows a large deviation from the DSMC result.
\begin{figure*}
\begin{center}
\includegraphics[width=0.48 \textwidth,height=0.36
\textwidth]{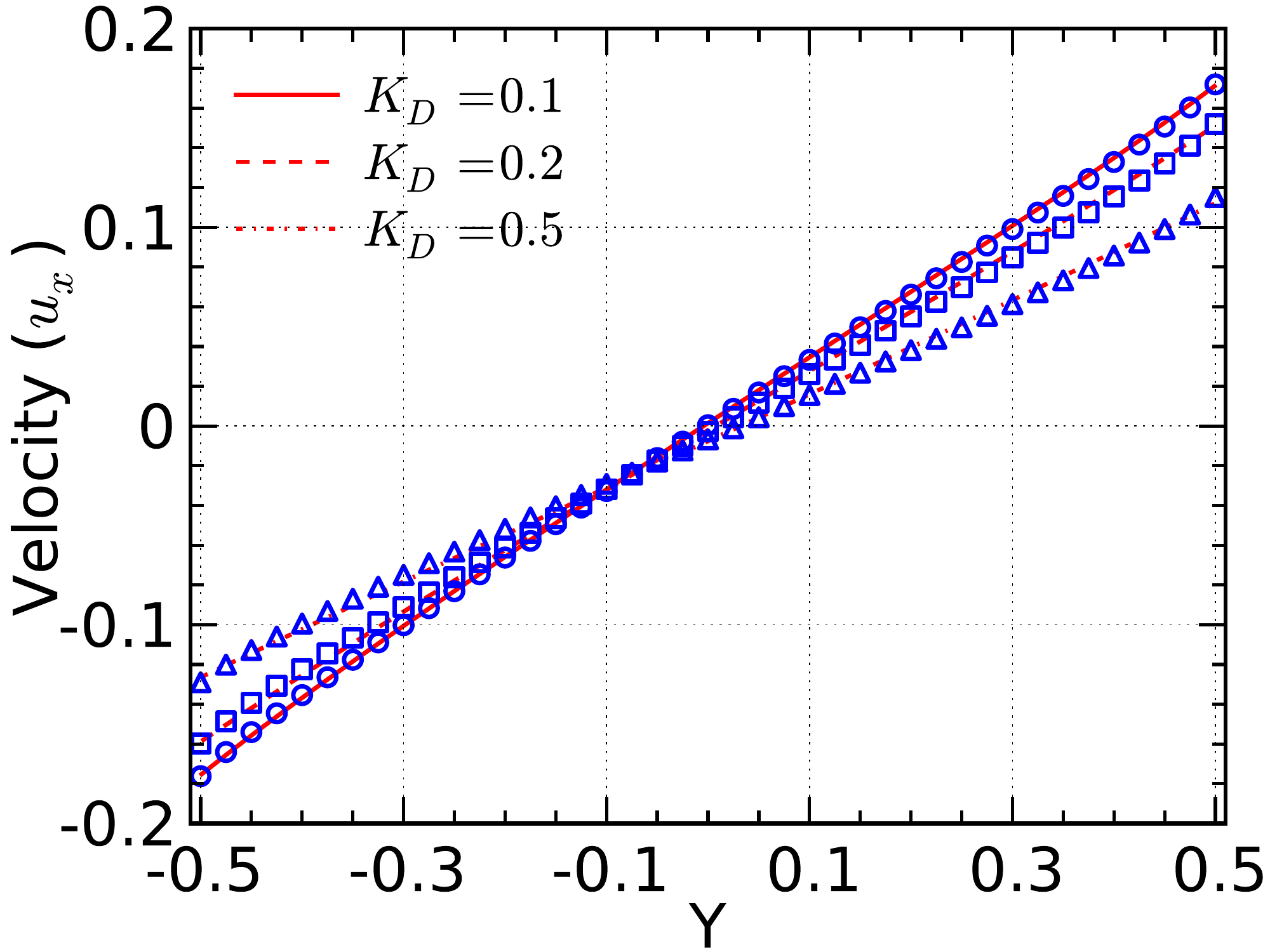}
\includegraphics[width=0.48 \textwidth,height=0.36
\textwidth]{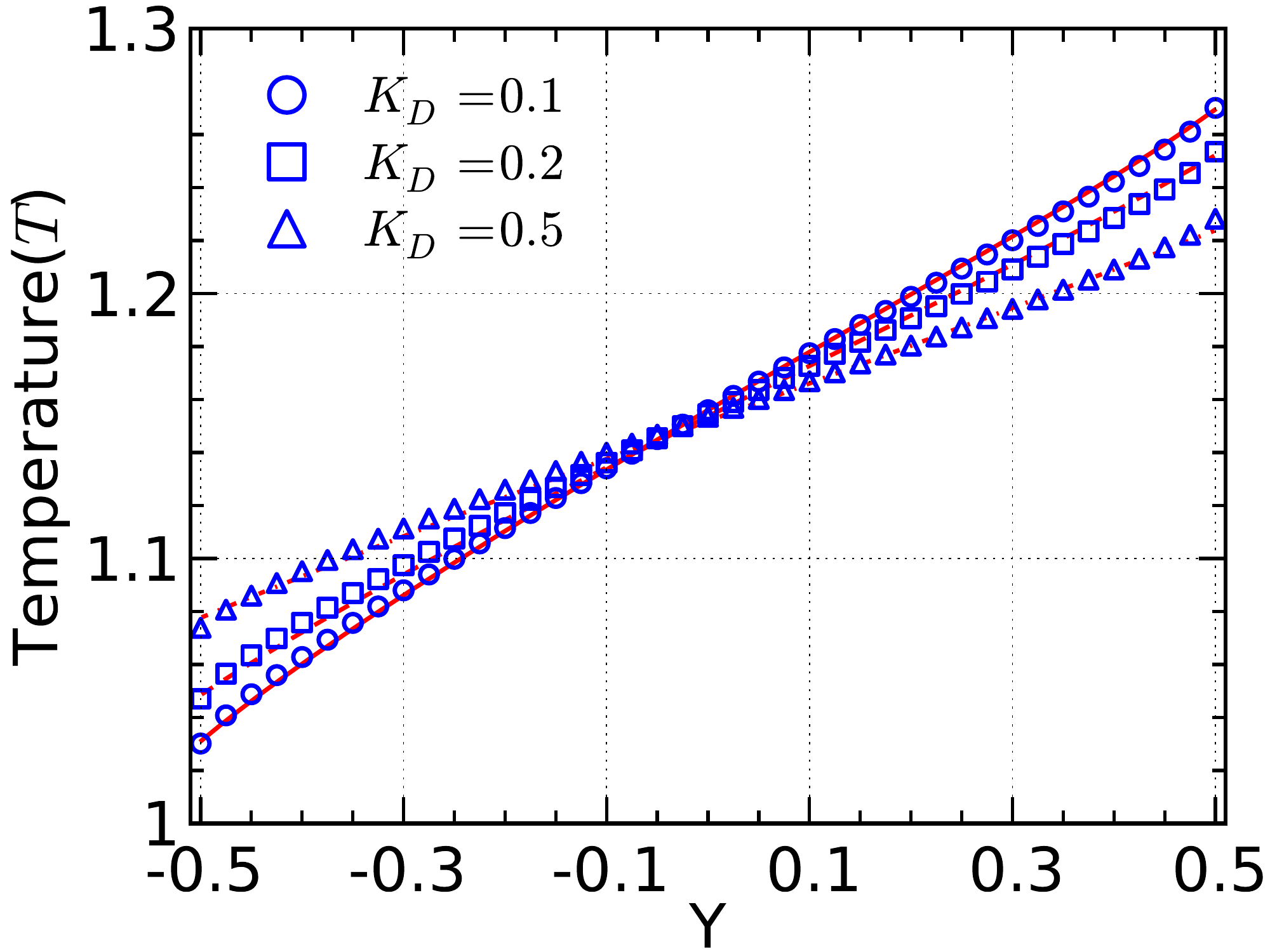}
\caption{Comparisons of the velocity and temperature profiles for
the combined Couette-Fourier flows. The wall velocities are
$u_w=\pm 0.2$ while the wall temperature difference is $0.3$.  The lines are the lattice ES-BGK model data
and the symbols are the results of the ES-BGK equation.\label{ccfut}}
\end{center}
\end{figure*}

While the comparisons with the  DSMC solutions 
for the temperature difference of $0.1$ have been shown in
Fig.~\ref{linearfouier}, a larger
temperature difference of $0.3$ is investigated for the
combined Couette-Fourier flow. The results are shown in
Fig.~\ref{ccfut} where excellent agreement is observed between the 
lattice ES-BGK model and the ES-BGK equation.

\section{Concluding remarks}

We have presented and validated a lattice Boltzmann model  using a 
systematic Hermite moment representation of the ES-BGK equation.  The 
resulting lattice ES-BGK model features an adjustable Prandtl number and may thus 
be more appropriate for use in coupled thermofluidic phenomena. 
This generic procedure may be applied to other kinetic model equations.

We have validated the lattice model for combined thermal and rarefaction 
effects by comparing its predictions with the DSMC and LVDSMC results as well 
as numerical solutions of the ES-BGK model. We find that, for 
a moderate 11th-order discrete velocity set, the proposed model 
can provide reasonable predictions for Couette and
force-driven Poiseuille flows for Knudsen numbers up to $0.5$. In addition, 
it is able to accurately predict heat
conduction in the slip-flow and early transition regimes. This finding 
extends to  unsteady problems provided the additional 
kinetic effects due to their time-dependence are also considered: for an  unsteady boundary 
heating problem reasonable agreement with LVDSMC simulations is observed for 
 Knudsen numbers up to $0.5$ and  Strouhal numbers up to $\pi \sqrt{2}/8$.

The solutions obtained by the lattice Boltzmann method are 
expected to approach the ``true'' solutions of the
ES-BGK equation as the number  of discrete velocities and the 
Hermite expansion order is increased. 
However, the moderate discrete velocity set used here already represents 
a reasonable compromise between computational efficiency and
modeling accuracy for flows with a range of Knudsen and Strouhal numbers. 
Although we have tested the model for rarefied gas thermal flows, this model, 
in principle, can be used for liquid thermal flows, which will be the subject of future work.

\end{document}